\pdfoutput=1
\documentclass[aps,prd,onecolumn,nofootinbib,amsmath,amssymb,floatfix,superscriptaddress,longbibliography,12pt]{revtex4-2}

\usepackage{graphicx}
\usepackage{slashed}
\usepackage{bm}
\usepackage{amsfonts}
\usepackage{mathrsfs}
\usepackage{xspace}
\usepackage{setspace}
\usepackage[colorlinks=true,linkcolor=blue,citecolor=blue,urlcolor=blue]{hyperref}

\newcommand{\gam}{\gamma}
\newcommand{\etap}{\eta^{\prime}}
\newcommand{\etapp}{\eta^{(\prime)}}
\newcommand{\pieta}{\pi^{0}\eta}
\newcommand{\pietap}{\pi^{0}\etap}
\newcommand{\pietapp}{\pi^{0}\etapp}
\newcommand{\etaeta}{\eta\eta}
\newcommand{\etaetap}{\eta\etap}
\newcommand{\etapetap}{\etap\etap}
\newcommand{\pizpiz}{\pi^{0}\pi^{0}}
\newcommand{\KsKs}{K^{0}_{S}K^{0}_{S}}
\newcommand{\bk}{\mathbf{k}}
\newcommand{\bb}{\mathbf{b}}
\newcommand{\kperp}{{\bk}_{\perp}}
\newcommand{\kperpi}[1]{{\bk}_{\perp #1}}
\newcommand{\xbar}{\bar{x}}
\newcommand{\ybar}{\bar{y}}
\newcommand{\mupi}{\mu_{\pi}}
\newcommand{\muq}{\mu_{\eta_{q}}}
\newcommand{\mus}{\mu_{\eta_{s}}}
\newcommand{\LamQCD}{\Lambda_{\mathrm{QCD}}}
\newcommand{\Hone}{\mathrm{H}_{0}^{(1)}}
\newcommand{\Jzero}{\mathrm{J}_{0}}
\newcommand{\Jone}{\mathrm{J}_{1}}
\newcommand{\Jtwo}{\mathrm{J}_{2}}
\newcommand{\Kzero}{\mathrm{K}_{0}}
\newcommand{\TildeT}{\widetilde{T}}

\begin{document}

\title{Twist-3 and flavor-singlet contributions to
       $\gam\gam\to\pi^{0}\etapp$ and $\etapp\etapp$
       in $k_{T}$ factorization}

\author{Cong Wang}
\email{wangj@hbnu.edu.cn}
\affiliation{College of Physics and Electronic Science, Hubei Normal University,
             Huangshi 435002, China}

\author{Jun-Kang He}
\email{hejk@hbnu.edu.cn}
\affiliation{College of Physics and Electronic Science, Hubei Normal University,
             Huangshi 435002, China}

\begin{abstract}
We study two-photon production of the pseudoscalar pairs $\pieta$, $\pietap$, $\etaeta$, $\etaetap$, and $\etapetap$ at center-of-mass energies
$2<Q<4~\mathrm{GeV}$ in $k_{T}$ factorization, including the chirally enhanced
twist-3 amplitudes and the two-gluon component of the $\etapp$.
The twist-3 amplitudes raise the hard cross sections by about an order of
magnitude in all five channels.  They bring $\etaeta$ close to the Belle
data, while $\pieta$ stays 3 to 5 times below and falls too slowly
with energy.  Varying the two-gluon distribution amplitude within its uncertainty changes the cross sections
by about 2\% in $\pieta$ and $\etaeta$ but by up to 24\% in the $\etap$
channels, making them its probe.  The hard and handbag
mechanisms each give nearly constant ratios $\sigma(\pieta)/\sigma(\pizpiz)$ and
$\sigma(\etaeta)/\sigma(\pizpiz)$, whereas below $3.2~\mathrm{GeV}$ the measured $\pieta$ ratio lies above
both predictions and the $\etaeta$ ratio below them.  Adding the handbag amplitude coherently, with its modulus
fixed by the pion and kaon channels and its phase fitted, reproduces the
$\pieta$ ratio up to $4~\mathrm{GeV}$ and the $\etaeta$ ratio below
$2.8~\mathrm{GeV}$, and steepens $\sigma(\pieta)$ to a $Q^{-n}$
dependence with $n=9.7\pm1.6$, against the measured $10.5\pm1.2\pm0.5$.  The
energy dependence of the cross section ratios can thus be understood as
the interference of two mechanisms of comparable size.
\end{abstract}

\maketitle

\section{Introduction}
\label{sec:intro}

Exclusive two-photon annihilation into a pseudoscalar meson pair is
among the cleanest testing grounds for perturbative QCD, since the
initial state is purely electromagnetic and, in the collinear
hard-scattering description, all nonperturbative input
resides in the light cone distribution amplitudes (LCDAs) of the
outgoing mesons.  At asymptotic energies the amplitude factorizes into a
hard kernel convoluted with two twist-2
LCDAs~\cite{Brodsky:1981rp,Lepage:1980fj}, which fixes
$d\sigma/d|\cos\theta^{*}|\propto\sin^{-4}\theta^{*}$ and
$\sigma\propto Q^{-6}$ for the charged pairs, where $Q$ is the two-photon
center-of-mass energy.  For a neutral pair the
$\sin^{-4}\theta^{*}$ term scales as $(e_{1}-e_{2})^{4}$, with $e_{1}$
and $e_{2}$ the charges of the quark flavors in the meson, and cancels
because the valence quark and antiquark are of the same flavor.  The
remaining leading-twist terms are small, so that formally
power-suppressed structures can exceed them, which turns the neutral
channels into a direct probe of those corrections.
Among them the two-parton twist-3 terms matter most at moderate energies.
They are chirally enhanced, the extra power of $1/Q$ coming with a mass
parameter set by the quark condensate rather than by a quark mass.  In a
companion study of $\gam\gam\to\pizpiz$ and $\KsKs$ they raised the
cross sections by close to an order of magnitude throughout
$Q\simeq2$--$4~\mathrm{GeV}$~\cite{He:2026bfv}.

The $\eta$ and $\etap$ carry a flavor-singlet component, which through
the $U(1)_{A}$ anomaly shapes their masses and mixing and couples them
directly to gluons~\cite{tHooft:1976rip}.  In two-photon pair production this
leaves three imprints.  Each meson is a mixture of a nonstrange and a
strange flavor state, described in the quark-flavor basis by a single
angle~\cite{Feldmann:1998vh,Feldmann:1998sh,Feldmann:1997vc}, so that
components with different charges add in one amplitude and the relative
sizes of the channels follow the mixing.  The normalization of the
twist-3 LCDAs follows from the anomalous Ward identity instead of the
Gell-Mann-Oakes-Renner relation, and because its direct evaluation
cancels strongly in the nonstrange sector, it is fixed by the
anomaly-free quark-mass part of the $\eta$-$\etap$ mass matrix.  And the singlet LCDA mixes under
evolution with a two-gluon LCDA~\cite{Kroll:2002nt,Agaev:2014wna}, so
that a $gg$ component of the $\etapp$ couples to the hard process already
at twist 2.  The channels $\gam\gam\to\pietapp$ and $\etapp\etapp$ thus
share one set of hard amplitudes, combined with different flavor weights,
and probe the mixing and the gluonic content in a large-angle exclusive
process, complementary to $B\to\etapp$
transitions~\cite{Charng:2006zj,Ball:2007hb},
$J/\psi\to\gam\etapp$~\cite{He:2019mpy}, $\etap$
photoproduction~\cite{Kroll:2021zss}, and the transition form
factors~\cite{Agaev:2014wna}.

Two of these channels, $\pieta$ and $\etaeta$, have been measured by
Belle~\cite{Uehara:2009cf,Uehara:2010mq}.  Above the resonance region
their cross sections fall as $Q^{-n}$ with $n=10.5\pm1.2\pm0.5$ for
$\pieta$ above $3.1~\mathrm{GeV}$ and $n=7.8\pm0.6\pm0.4$ for $\etaeta$
between $2.4$ and $3.3~\mathrm{GeV}$, against $8.0\pm0.5\pm0.4$ for
$\pizpiz$~\cite{Uehara:2009cka}.  The cross section ratios
$\sigma(\pieta)/\sigma(\pizpiz)$ and $\sigma(\etaeta)/\sigma(\pizpiz)$ are the
sharper test, since normalizations and much of the systematics cancel.
Belle quoted their averages, $0.48\pm0.05\pm0.04$ for $\pieta$ and
$0.37\pm0.02\pm0.03$ for $\etaeta$, and found that the leading-term
counting with $\theta_{P}=-18^{\circ}$ and $f_{\eta}=f_{\pi}$ reproduces
the first and overshoots the second~\cite{Uehara:2009cf,Uehara:2010mq}.
The ratios are, however, not constant in
energy~\cite{Uehara:2009cka,Uehara:2009cf,Uehara:2010mq}.  In the
central angular range the $\pieta$ ratio falls from about $0.75$ below
$3.2~\mathrm{GeV}$ to about $0.45$ near $4~\mathrm{GeV}$, and the $\etaeta$
ratio from about $0.37$ to about $0.25$ above $2.8~\mathrm{GeV}$,
whereas the leading term fixes the ratios through the quark charges and
the mixing alone.  The angular distribution of $\etaeta$, moreover,
prefers $\sin^{-6}\theta^{*}$ above $3~\mathrm{GeV}$, unlike those of
$\pizpiz$ and $\pieta$~\cite{Uehara:2010mq}.  The
channels $\pietap$, $\etaetap$, and $\etapetap$ remain unmeasured.

Theoretically these final states have received far less attention than
their pionic and kaonic counterparts.  Chernyak and
Eidelman~\cite{Chernyak:2006dk,Chernyak:2009dj,Chernyak:2014wra} analyzed
them within leading-term QCD and pointed to odderon
exchange~\cite{Ginzburg:1992mi}, a gluonic mechanism specific to
charge-suppressed final states, whose size depends strongly on the LCDAs
and is reduced further by $\eta$-$\etap$ mixing in $\pieta$.  In the soft
handbag mechanism of Diehl, Kroll, and Vogt~\cite{Diehl:2001fv} the two photons couple to a single
quark line and the pair is formed through a soft annihilation form
factor.  With two such form factors fitted to the measured channels,
Ref.~\cite{Diehl:2009yi} predicted
$\sigma(\etaeta)/\sigma(\pizpiz)\approx0.3$, close to the value Belle
later measured.  The soft and the hard descriptions are not alternatives
at these energies.  In the soft-collinear effective theory analysis of
$\gam\gam\to\pi\pi$~\cite{Kivel:2013fja} a consistent factorization at
subleading power keeps the soft-overlap contribution together with the
chirally enhanced twist-3 one.  Below the perturbative region the $\pieta$
system has been treated with dispersive and coupled-channel
methods~\cite{Danilkin:2017lyn,Lu:2020qeo}, and BESIII has measured
$\gam\gam\to\pieta$ together with $\pizpiz$ and $K^{+}K^{-}$ and
analyzed the three channels in a coupled-channel partial wave
analysis~\cite{Lellmann:2025qii}.

For the $\etapp$ channels neither the $gg$ component nor the soft
contribution has yet been combined with the twist-3 hard amplitudes.  We
compute the hard amplitudes in $k_{T}$
factorization~\cite{Botts:1989kf,Li:1992nu,Jakob:1993iw}, which has been
applied to two-photon meson-pair production in
Refs.~\cite{Coriano:1998ge,Hsieh:2004ee}.  Keeping the parton transverse
momentum in the propagators removes the end point singularities of the
twist-3 convolution, and the Sudakov factor and the threshold-resummed
jet function~\cite{Li:1998is,Kurimoto:2001zj,Li:2001ay} resum the
associated logarithms and suppress the end point region, in which the
smallest partonic virtuality in the angular acceptance approaches
$\LamQCD^{2}$ at intermediate energies.  With the chirally enhanced
twist-3 LCDAs the framework describes the charged pion and kaon
pairs~\cite{Wang:2015mod,Wang:2019trq} and brings the neutral ones much
closer to the data~\cite{He:2026bfv}.  Here the twist-2
sector contains the octet-singlet decomposition with the quark-gluon
mixing of the singlet and two-gluon LCDAs, and the twist-3 sector the
nonstrange and strange components with their own chiral masses.  In the
neutral pion and kaon channels the soft handbag amplitude turned out to be
comparable to the hard one in the Belle energy range, and the two added
coherently largely reproduced the measured energy dependence of the
neutral-to-charged ratios~\cite{He:2026bfv}.  We follow the same
approach here, taking the modulus of the soft amplitude from those
channels and fitting its phase to the measured $\pieta$ and $\etaeta$
cross section ratios.

We address three questions, namely how far the hard mechanism with the
chirally enhanced twist-3 amplitudes accounts for the measured cross
sections, which channels are sensitive to the two-gluon component, and
what produces the energy dependence of the ratios to $\pizpiz$.  To this
end we compute the cross sections and angular distributions of all five
channels with their theoretical uncertainties, compare the two measured
ones with the Belle data, predict the other three, and analyze the ratios
to $\pizpiz$ in the coherent sum of the soft and hard amplitudes.
Section~\ref{sec:formalism} sets up the framework,
Sec.~\ref{sec:numerics} collects the inputs and presents the results, and
Sec.~\ref{sec:summary} summarizes.

\section{Theoretical framework}
\label{sec:formalism}

\subsection{Kinematics and cross section}
\label{subsec:kinematics}

We work in the center-of-mass frame of the two-photon system and let the
meson $P_{1}(p_{3})$ define the positive $z$ axis.  With light cone
coordinates $A^{\mu}=(A^{+},A^{-},\mathbf{A}_{\perp})$ and
$A^{\pm}=A^{0}\pm A^{3}$, the external momenta of
$\gam(p_{1},\lambda_{1})\gam(p_{2},\lambda_{2})\to
P_{1}(p_{3})P_{2}(p_{4})$ read
\begin{equation}
\label{eq:kinmomenta}
\begin{aligned}
p_{1}^{\mu}&=(2\omega c^{2},\,2\omega s^{2},\,\mathbf{p}_{\perp}),
&\quad
p_{2}^{\mu}&=(2\omega s^{2},\,2\omega c^{2},\,-\mathbf{p}_{\perp}),\\
p_{3}^{\mu}&=(2\omega,\,0,\,\mathbf{0}_{\perp}),
&\quad
p_{4}^{\mu}&=(0,\,2\omega,\,\mathbf{0}_{\perp}),
\end{aligned}
\end{equation}
where $\omega=Q/2$ is the half center-of-mass energy and
$\mathbf{p}_{\perp}=(2\omega sc,0)$.  The abbreviations for the
scattering angle $\theta^{*}$ of $P_{1}$ relative to $\gam(p_{1})$ are
\begin{equation}
\label{eq:sc}
s=\sin\frac{\theta^{*}}{2},
\qquad
c=\cos\frac{\theta^{*}}{2}.
\end{equation}
The photon polarization vectors are those of Ref.~\cite{He:2026bfv}, and
the Mandelstam invariants reduce to $t=-Q^{2}s^{2}$ and $u=-Q^{2}c^{2}$
with $t+u=-Q^{2}$.

The meson masses enter only through power corrections of relative order
$m_{P}^{2}/Q^{2}$.  We neglect them consistently, in the hard kernels as
well as in Eq.~\eqref{eq:kinmomenta}, as in the treatments of the pion
and kaon channels~\cite{Wang:2015mod,Wang:2019trq,He:2026bfv} and of the
handbag contribution~\cite{Diehl:2001fv,Diehl:2009yi}.  These corrections
are largest for the heaviest final state, $\etapetap$, close to its
threshold.  Averaging over the initial photon polarizations, the
differential cross section reads
\begin{equation}
\label{eq:xsec}
\frac{d\sigma}{d\cos\theta^{*}}
=\frac{1}{32\pi Q^{2}\,\mathcal{S}}\,\frac{1}{4}
\sum_{\lambda_{1},\lambda_{2}=\pm}
\bigl|\mathcal{M}_{\lambda_{1}\lambda_{2}}\bigr|^{2},
\end{equation}
with the statistical factor
\begin{equation}
\label{eq:statfac}
\mathcal{S}=
\begin{cases}
1, & P_{1}\neq P_{2}\quad(\pieta,\ \pietap,\ \etaetap),\\[2pt]
2, & P_{1}=P_{2}\quad(\etaeta,\ \etapetap),
\end{cases}
\end{equation}
which accounts for the identical particles in the last two channels.
Parity and the $C$ invariance of the two-photon state leave two
independent helicity amplitudes, the equal-helicity amplitude
$\mathcal{M}_{++}=\mathcal{M}_{--}$ and the opposite-helicity amplitude
$\mathcal{M}_{+-}=\mathcal{M}_{-+}$, so that the sum in
Eq.~\eqref{eq:xsec} reduces to
$2(|\mathcal{M}_{++}|^{2}+|\mathcal{M}_{+-}|^{2})$.  We keep the two
separate throughout.  They have different angular dependences and
receive different hadronic contributions, the two-gluon and the soft
handbag amplitudes entering $\mathcal{M}_{+-}$ alone.

The reliability of the perturbative treatment is controlled not by
$Q^{2}$ but by the smaller invariants $|t|$ and $|u|$, which set the
virtualities of the internal propagators.  As
$\min(|t|,|u|)=Q^{2}\min(s^{2},c^{2})$ becomes parametrically small
toward the forward and backward directions, we restrict the angular
range to $|\cos\theta^{*}|<0.6$, for which $\min(|t|,|u|)>0.2\,Q^{2}$
stays safely above the soft scale.  It is also the range used by Belle
for the charged pion and kaon pairs and for
$\KsKs$~\cite{Nakazawa:2004gu,Chen:2006gy}, and the published $\pieta$
and $\etaeta$ differential cross sections extend over
it~\cite{Uehara:2009cf,Uehara:2010mq}.

\subsection{Light cone distribution amplitudes}
\label{subsec:lcda}

The nonperturbative input is carried by the LCDAs, defined by
gauge-invariant bilocal matrix elements between the meson and the
vacuum.  At the two-parton level there is one LCDA of twist 2, from
the axial-vector projection, and there are two of twist 3, from the
pseudoscalar and pseudotensor projections.  The genuine three-parton
$q\bar qg$ LCDA of twist 3 is tied to these by the QCD equations of
motion and survives in their shapes through the ratio
$\eta_{3P}=f_{3P}/(f_{P}\mu_{P})$ of its coupling $f_{3P}$ to the
normalization $f_{P}\mu_{P}$ of the two-parton twist-3 LCDAs of the
meson $P$~\cite{Braun:1989iv,Ball:1998je,Ball:2006wn}.  Its
explicit contribution to the hard scattering is dropped, as in the
analyses of the pion and kaon
channels~\cite{Wang:2015mod,Wang:2019trq,He:2026bfv}.  With
$\eta_{3P}\sim10^{-2}$ the neglected terms lie 2 orders of magnitude
below the two-parton twist-3 terms that are kept, and the present
results remain directly comparable with those of the pion and kaon
channels.

For the $\pi^{0}$ we follow Ref.~\cite{He:2026bfv}.  With $z=z_{1}-z_{2}$
lightlike, the three matrix elements
read~\cite{Braun:1989iv,Ball:1998je,Ball:2006wn}
\begin{align}
\langle\pi^{0}(p)|\bar q(z_{1})\gamma_{\mu}\gamma_{5}q(z_{2})|0\rangle
 &=-i f_{\pi}\,p_{\mu}\!\int_{0}^{1}\!\!dx\,
   e^{i(xp\cdot z_{1}+\xbar p\cdot z_{2})}\,\phi_{\pi}^{\,\pi}(x,\mu),
   \nonumber\\
\langle\pi^{0}(p)|\bar q(z_{1})i\gamma_{5}q(z_{2})|0\rangle
 &=f_{\pi}\mupi\!\int_{0}^{1}\!\!dx\,
   e^{i(xp\cdot z_{1}+\xbar p\cdot z_{2})}\,\phi_{\pi}^{p}(x,\mu),
   \nonumber\\
\langle\pi^{0}(p)|\bar q(z_{1})i\gamma_{5}\sigma_{\mu\nu}q(z_{2})|0\rangle
 &=-f_{\pi}\mupi\bigl(p_{\mu}z_{\nu}-p_{\nu}z_{\mu}\bigr)
   \!\int_{0}^{1}\!\!dx\,e^{i(xp\cdot z_{1}+\xbar p\cdot z_{2})}\,
   \frac{\phi_{\pi}^{\sigma}(x,\mu)}{6},
\label{eq:pionME}
\end{align}
where $\bar q\,\Gamma\,q=(\bar u\,\Gamma\,u-\bar d\,\Gamma\,d)/\sqrt2$ is
the isovector bilinear for each Dirac structure $\Gamma$.  The
superscript on $\phi$ labels the Dirac structure, $\phi_{\pi}^{\,\pi}$
being the twist-2 LCDA and $\phi_{\pi}^{p}$ and $\phi_{\pi}^{\sigma}$
the twist-3 ones.  All three are normalized to unity, so that the size of
each matrix element is carried by its prefactor, $f_{\pi}$ at twist 2 and
$f_{\pi}\mupi$ at twist 3.  The twist-3
normalization follows from the divergence of the axial current, which
for a nonsinglet channel contains only the quark-mass term, so that the
Gell-Mann-Oakes-Renner relation~\cite{GellMann:1968rz,Leutwyler:1996qg}
gives the chiral mass
\begin{equation}
\label{eq:mupi}
\mupi(\mu)=\frac{m_{\pi}^{2}}{m_{u}(\mu)+m_{d}(\mu)}
=-\frac{2\langle\bar qq\rangle(\mu)}{f_{\pi}^{2}}+\mathcal{O}(m_{q}).
\end{equation}
The product $f_{\pi}\mupi$ thus measures the quark condensate rather than
a quark mass.  The twist-3 contributions are formally suppressed by
$\mupi/Q$, but with a coefficient of order $\LamQCD$, which is the origin
of the chiral enhancement.

For the $\etapp$ we use the Feldmann-Kroll-Stech mixing
scheme~\cite{Feldmann:1998vh,Feldmann:1998sh}, in which the physical
mesons are orthogonal rotations of the flavor states
$|\eta_{q}\rangle=(|u\bar u\rangle+|d\bar d\rangle)/\sqrt2$ and
$|\eta_{s}\rangle=|s\bar s\rangle$ through a single angle $\phi$
\begin{equation}
\label{eq:statemix}
\begin{pmatrix} |\eta\rangle\\ |\etap\rangle\end{pmatrix}
=\begin{pmatrix} \cos\phi & -\sin\phi\\
                 \sin\phi & \cos\phi\end{pmatrix}
 \begin{pmatrix} |\eta_{q}\rangle\\ |\eta_{s}\rangle\end{pmatrix}.
\end{equation}
The relevant bilinears are
$\bar q\,\Gamma\,q=(\bar u\,\Gamma\,u+\bar d\,\Gamma\,d)/\sqrt2$ for the
nonstrange component and $\bar s\,\Gamma\,s$ for the strange one, and the
twist-2 matrix element of the component $r=q,s$ of the physical state
$P=\eta,\etap$ reads
\begin{equation}
\label{eq:etaME2}
\langle P(p)|\bar q(z_{1})\gamma_{\mu}\gamma_{5}q(z_{2})|0\rangle
=-i f_{P}^{r}\,p_{\mu}\!\int_{0}^{1}\!\!dx\,
  e^{i(xp\cdot z_{1}+\xbar p\cdot z_{2})}\,\phi_{r}^{\,\pi}(x,\mu),
\end{equation}
the first line of Eq.~\eqref{eq:pionME} with $f_{\pi}$ replaced by the
flavor decay constant $f_{P}^{r}$ and with $\phi_{r}^{\,\pi}$ the twist-2
LCDA of the flavor state $\eta_{r}$.  Since the physical states are
the rotations of Eq.~\eqref{eq:statemix}, the four decay constants follow the same pattern
\begin{equation}
\label{eq:fmix}
\begin{pmatrix} f_{\eta}^{q} & f_{\eta}^{s}\\
                f_{\etap}^{q} & f_{\etap}^{s}\end{pmatrix}
=\begin{pmatrix} \cos\phi & -\sin\phi\\
                 \sin\phi & \cos\phi\end{pmatrix}
 \begin{pmatrix} f_{q} & 0\\ 0 & f_{s}\end{pmatrix},
\end{equation}
with $f_{q}$ and $f_{s}$ the couplings of the flavor states.  Two
couplings and one angle thus replace the four decay constants, and the
negative sign of $f_{\eta}^{s}$ is fixed by the rotation.  The mixing
resides in the normalizations alone, while the shapes belong to the
flavor states, so that one and the same $\phi_{q}^{\,\pi}$, and one and
the same $\phi_{s}^{\,\pi}$, serve both the $\eta$ and the
$\etap$~\cite{Feldmann:1998vh,Agaev:2014wna}.  This holds at a reference
scale $\mu_{0}\simeq1~\mathrm{GeV}$ and is modified above it by
evolution.

At twist 3 the flavor-singlet nature of the $\etapp$ enters through the
twist-3 normalizations.  In $\gam\gam\to PP$ the leading twist-3
effect is a double chirality flip, one in each meson, so that no quark
mass appears in the hard part and the amplitude carries the product of
the two normalizations.  As for the $\pi^{0}$, they are fixed by the
divergence of the axial current, which for a single light flavor now carries the anomaly
\begin{equation}
\label{eq:AWI}
\partial^{\mu}\bigl(\bar q\gamma_{\mu}\gamma_{5}q\bigr)
=2m_{q}\,\bar q\,i\gamma_{5}q
+\frac{\alpha_{s}}{4\pi}G^{A}_{\mu\nu}\widetilde G^{A,\mu\nu},
\end{equation}
where $G^{A}_{\mu\nu}$ is the gluon field strength,
$\widetilde G^{A,\mu\nu}=\tfrac12\epsilon^{\mu\nu\alpha\beta}
G^{A}_{\alpha\beta}$ its dual and $\epsilon_{0123}=1$.  Taking the matrix
element of Eq.~\eqref{eq:AWI} and of its strange counterpart between the
vacuum and a physical state gives~\cite{Feldmann:1998vh,Beneke:2002jn}
\begin{equation}
\label{eq:AWIrel}
f_{P}^{q}m_{P}^{2}=h_{P}^{q}+\sqrt2\,a_{P},
\qquad
f_{P}^{s}m_{P}^{2}=h_{P}^{s}+a_{P},
\end{equation}
in which $a_{P}=\langle0|(\alpha_{s}/4\pi)G^{A}_{\mu\nu}
\widetilde G^{A,\mu\nu}|P\rangle$ is the anomaly matrix element, the
factor $\sqrt2$ counting its contribution to the isoscalar current, and
$h_{P}^{r}=2m_{r}\langle0|\bar r\,i\gamma_{5}r|P\rangle$ are the
scale-independent pseudoscalar densities, with $m_{q}=(m_{u}+m_{d})/2$
and $\bar r\,i\gamma_{5}r$ the nonstrange or strange bilinear defined
above.  The meson mass thus splits into a quark-mass piece, measured by
$h_{P}^{r}$, and a gluonic piece generated by the anomaly.  The Okubo-Zweig-Iizuka (OZI) rule underlying the mixing scheme applies to the densities as to the decay constants
\begin{equation}
\label{eq:hmix}
\begin{pmatrix} h_{\eta}^{q} & h_{\eta}^{s}\\
                h_{\etap}^{q} & h_{\etap}^{s}\end{pmatrix}
=\begin{pmatrix} \cos\phi & -\sin\phi\\
                 \sin\phi & \cos\phi\end{pmatrix}
 \begin{pmatrix} h_{q} & 0\\ 0 & h_{s}\end{pmatrix},
\end{equation}
and the twist-3 LCDAs are defined by the light cone generalization of the densities
\begin{align}
2m_{r}\langle P(p)|\bar q(z_{1})i\gamma_{5}q(z_{2})|0\rangle
 &=h_{P}^{r}\!\int_{0}^{1}\!\!dx\,
   e^{i(xp\cdot z_{1}+\xbar p\cdot z_{2})}\,\phi_{r}^{p}(x,\mu),
   \nonumber\\
2m_{r}\langle P(p)|\bar q(z_{1})i\gamma_{5}\sigma_{\mu\nu}q(z_{2})|0\rangle
 &=-h_{P}^{r}\bigl(p_{\mu}z_{\nu}-p_{\nu}z_{\mu}\bigr)
   \!\int_{0}^{1}\!\!dx\,e^{i(xp\cdot z_{1}+\xbar p\cdot z_{2})}\,
   \frac{\phi_{r}^{\sigma}(x,\mu)}{6},
\label{eq:etaME3}
\end{align}
with $\phi_{r}^{p}$ and $\phi_{r}^{\sigma}$ the twist-3 LCDAs of the
flavor state $\eta_{r}$, normalized to unity and, as at twist 2, shared
by the $\eta$ and the $\etap$.  Trading the densities for
chiral masses, the analogs of $\mupi$,
\begin{equation}
\label{eq:mudef}
\muq=\frac{h_{q}}{2m_{q}f_{q}},
\qquad
\mus=\frac{h_{s}}{2m_{s}f_{s}},
\end{equation}
brings Eq.~\eqref{eq:etaME3} to the form of the last two lines of
Eq.~\eqref{eq:pionME} with $f_{\pi}\mupi$ replaced by $f_{P}^{q}\muq$ or
$f_{P}^{s}\mus$.  The
twist-3 amplitudes therefore enter through $\mupi\muq$ in $\pietapp$ and
through $\muq^{2}$, $\muq\mus$, or $\mus^{2}$ in $\etapp\etapp$, and the
values of the two scales have to be fixed.

Solving Eq.~\eqref{eq:AWIrel} within the mixing pattern of
Eq.~\eqref{eq:fmix} gives the familiar
expressions~\cite{Beneke:2002jn,Ball:2007hb,Agaev:2014wna}
\begin{align}
\label{eq:hqhs}
h_{q}&=f_{q}\bigl(m_{\eta}^{2}\cos^{2}\phi+m_{\etap}^{2}\sin^{2}\phi\bigr)
      -\sqrt2 f_{s}\bigl(m_{\etap}^{2}-m_{\eta}^{2}\bigr)\sin\phi\cos\phi,
      \nonumber\\
h_{s}&=f_{s}\bigl(m_{\eta}^{2}\sin^{2}\phi+m_{\etap}^{2}\cos^{2}\phi\bigr)
      -\frac{f_{q}}{\sqrt2}\bigl(m_{\etap}^{2}-m_{\eta}^{2}\bigr)
       \sin\phi\cos\phi.
\end{align}
For $h_{s}$ the first term exceeds the second by a factor of about 4 and
the result is stable.  For $h_{q}$ the two terms cancel to about 4\%,
leaving a value some 30 times below $h_{s}$ whose uncertainty
exceeds its central value once the errors on $f_{q}$, $f_{s}$, and $\phi$
are propagated.  The cancellation has a physical origin.  Without the
anomaly $\eta_{q}$ would be a Goldstone boson degenerate with the pion,
and the anomaly matrix element removes almost all of
$m_{P}^{2}f_{P}^{q}$, leaving only the small Goldstone piece.
Equation~\eqref{eq:hqhs}, exact as an identity, is therefore a poor
numerical definition of $\muq$, and the smallness of $h_{q}$ says nothing
about the size of the twist-3 amplitude itself.

The literature handles this in different ways.  In Ref.~\cite{Charng:2006zj} Eq.~\eqref{eq:hqhs} is used for both flavors, in Ref.~\cite{Ball:2007hb} the $\eta_{q}$ normalization is replaced by its leading-order chiral value, in Ref.~\cite{Huang:2026hvs} an $SU(3)$-symmetric prescription is adopted, and in Ref.~\cite{Agaev:2014wna} the $\eta_{q}$ twist-3 amplitudes are dropped.  The last
choice is appropriate for the transition form factor
$\gam^{*}\gam\to\etapp$, where the single chirality flip has to be
compensated by a light-quark mass insertion in the hard part, so that the
contribution is proportional to $h_{q}$ itself and is genuinely
suppressed.  It does not carry over to $\gam\gam\to PP$, where the double
flip involves no quark mass and the $\eta_{q}$ twist-3 amplitudes enter
at full strength.

A stable prescription follows from the flavor basis.  By
Eq.~\eqref{eq:AWIrel} the mass-squared matrix of the $\eta$-$\etap$
system splits into an anomaly part and a quark-mass part that the OZI
rule makes diagonal, with entries $m_{qq}^{2}=h_{q}/f_{q}$ and
$m_{ss}^{2}=h_{s}/f_{s}$~\cite{Feldmann:1998vh}.  Being free of the
anomaly, these are governed by chiral symmetry breaking alone and are
given by the pion and kaon masses~\cite{Feldmann:1998vh,Huang:2026hvs},
$m_{qq}^{2}=m_{\pi}^{2}$ up to OZI-violating corrections, since $\eta_{q}$
shares the quark content of the pion and differs from it only in
isospin, and $m_{ss}^{2}=2m_{K}^{2}-m_{\pi}^{2}$ at leading chiral order.
Inserting these into Eq.~\eqref{eq:mudef} gives
\begin{equation}
\label{eq:muqs}
\muq=\frac{m_{\pi}^{2}}{2m_{q}}=\mupi,
\qquad
\mus=\frac{2m_{K}^{2}-m_{\pi}^{2}}{2m_{s}}.
\end{equation}
For $\eta_{q}$ this coincides with the replacement
$h_{q}\to f_{q}m_{\pi}^{2}$ of Ref.~\cite{Ball:2007hb}.  Both scales
inherit only the small uncertainties of the quark masses and treat the
two flavor components on the same footing, so that the ratio $\mus/\muq$,
which sets the relative weight of the two flavor sectors in the twist-3
amplitude, is not distorted by a cancellation that affects $h_{q}$ alone.

The $\pi^{0}$, the $\eta_{q}$, and the $\eta_{s}$ are all flavor
neutral and $C$ even, so their LCDAs are symmetric under
$x\leftrightarrow\xbar$, odd Gegenbauer moments are absent and no
end point logarithms of the kaon type appear.  The conformal expansions
therefore have a common structure, which we write once with
$r=\pi,q,s$ labeling in turn the $\pi^{0}$ and the two flavor states
$\eta_{q}$ and $\eta_{s}$ of the $\etapp$.  With $\xi=2x-1$, and truncating after
the second nontrivial moment,
\begin{align}
\phi_{r}^{\,\pi}(x,\mu)&=6x\xbar\Bigl[1+a_{2}^{r}(\mu)C_{2}^{3/2}(\xi)
   +a_{4}^{r}(\mu)C_{4}^{3/2}(\xi)\Bigr],\nonumber\\
\phi_{r}^{p}(x,\mu)&=1+\bigl[30\eta_{3r}-\tfrac52\rho_{r}^{2}\bigr]
   C_{2}^{1/2}(\xi)
   -\bigl[3\eta_{3r}\omega_{3r}+\tfrac{27}{20}\rho_{r}^{2}
   +\tfrac{81}{10}\rho_{r}^{2}a_{2}^{r}\bigr]C_{4}^{1/2}(\xi),\nonumber\\
\phi_{r}^{\sigma}(x,\mu)&=6x\xbar\Bigl\{1+\bigl[5\eta_{3r}
   -\tfrac12\eta_{3r}\omega_{3r}-\tfrac{7}{20}\rho_{r}^{2}
   -\tfrac35\rho_{r}^{2}a_{2}^{r}\bigr]C_{2}^{3/2}(\xi)\Bigr\},
\label{eq:DAexp}
\end{align}
where the Gegenbauer polynomials entering the twist-3 LCDAs are
$C_{2}^{1/2}(\xi)=\tfrac12(3\xi^{2}-1)$ and
$C_{4}^{1/2}(\xi)=\tfrac18(35\xi^{4}-30\xi^{2}+3)$.
Here $\eta_{3r}$ is the ratio defined above, taken for
the flavor state $r$, and $\omega_{3r}$ the gluon
momentum-fraction parameter, both taken over from the pion
analyses~\cite{Ball:1998je,Ball:2006wn}, while the quark-mass
parameters are
\begin{equation}
\label{eq:rho}
\rho_{\pi}^{2}=\rho_{q}^{2}=\frac{(m_{u}+m_{d})^{2}}{m_{\pi}^{2}},
\qquad
\rho_{s}^{2}=\frac{4m_{s}^{2}}{2m_{K}^{2}-m_{\pi}^{2}}.
\end{equation}
The two behave very differently.  Whereas $\rho_{\pi}^{2}\sim10^{-3}$ is
numerically irrelevant and is kept only for consistency,
$\rho_{s}^{2}\simeq0.07$ is not.  In the second line of
Eq.~\eqref{eq:DAexp} the term $-\tfrac52\rho_{r}^{2}$ is then comparable
in size to the three-parton term $30\eta_{3r}$ and of opposite sign, so
the strange twist-3 LCDA departs visibly from its asymptotic form
and the mass correction has to be retained.

The flavor-singlet component of the $\etapp$ also couples to a
two-gluon state, which has no counterpart in the $\pi^{0}$.  Since
the two gluons attach directly to a quark line in the hard scattering,
the form needed here is the twist-2 two-gluon matrix element with open
gluon fields given in Ref.~\cite{Ball:2007hb}.  With $z=z_{1}-z_{2}$ and $A$, $B$ adjoint
color indices,
\begin{equation}
\label{eq:gluonME}
\langle P(p)|A^{A}_{\alpha}(z_{1})A^{B}_{\beta}(z_{2})|0\rangle
=\frac{\delta^{AB}}{N_{c}^{2}-1}\,\frac{C_{F}}{4\sqrt3}\,f_{P}^{1}\,
 \epsilon_{\alpha\beta\rho\sigma}\,\frac{z^{\rho}p^{\sigma}}{p\cdot z}
 \int_{0}^{1}\!\!dx\,e^{i(xp\cdot z_{1}+\xbar p\cdot z_{2})}\,
 \frac{\phi_{g}(x,\mu)}{x\xbar},
\end{equation}
which is equivalent to the gauge-invariant definition in terms of
$G\widetilde G$ used in Refs.~\cite{Kroll:2002nt,Agaev:2014wna}.  The
singlet decay constant follows from the same mixing parameters
\begin{equation}
\label{eq:f1def}
f_{P}^{1}=\frac{\sqrt2f_{P}^{q}+f_{P}^{s}}{\sqrt3},
\end{equation}
and runs through the two-loop $U(1)_{A}$ anomalous
dimension~\cite{Kodaira:1979pa,Agaev:2014wna}
\begin{equation}
\label{eq:f1run}
f_{P}^{1}(\mu)=f_{P}^{1}(\mu_{0})
\left\{1+\frac{2n_{f}}{\pi\beta_{0}}
\bigl[\alpha_{s}(\mu)-\alpha_{s}(\mu_{0})\bigr]\right\},
\end{equation}
with $\beta_{0}=\tfrac{11}{3}N_{c}-\tfrac23n_{f}$.  This running is a
genuine singlet effect.  The octet combination
$f_{P}^{8}=(f_{P}^{q}-\sqrt2f_{P}^{s})/\sqrt3$ belongs to an axial current
that is conserved in the chiral limit and does not run, whereas the
anomaly of Eq.~\eqref{eq:AWI} survives only in the singlet current.  Charge conjugation makes $\phi_{g}$
antisymmetric, $\phi_{g}(x)=-\phi_{g}(\xbar)$, so that it integrates to
zero and, unlike the quark LCDAs of Eq.~\eqref{eq:DAexp}, cannot be
normalized to unity.  Its size is carried entirely by the coefficients of
its conformal expansion.  There is no two-gluon counterpart of the
chirally enhanced twist-3 LCDAs, which flip the quark chirality.  The conformal expansion runs over Gegenbauer polynomials
$C_{n-1}^{5/2}$ of odd degree and, truncated at $n=2$, reads
\begin{equation}
\label{eq:gluonDA}
\phi_{g}(x,\mu)=x^{2}\xbar^{2}\,a_{2}^{g}(\mu)\,C_{1}^{5/2}(2x-1)+\cdots ,
\end{equation}
where $C_{1}^{5/2}(\xi)=5\xi$.  As with the quark LCDAs the mixing
scheme identifies $\phi_{g}$ for the $\eta$ and the $\etap$, the whole
difference between the two residing in $f_{P}^{1}$.  The coefficient
$a_{2}^{g}$ is defined as in Ref.~\cite{Ball:2007hb}, where it coincides
with the coefficient quoted in Ref.~\cite{Kroll:2002nt}.
The expansion of Ref.~\cite{Agaev:2014wna} carries an extra factor
$30$, so that its coefficient equals $a_{2}^{g}/30$.

Two features of Eq.~\eqref{eq:gluonDA} matter for what follows.  Having
no zeroth moment, the two-gluon LCDA cannot rescale the quark
contribution but enters only through an interference with it, which changes
sign under $x\leftrightarrow\xbar$.  The prefactor $x^{2}\xbar^{2}$ makes
$\phi_{g}$ vanish quadratically at both end points, one power faster than
the quark LCDAs, so the gluonic convolution is end point safe and
needs no transverse-momentum regularization, which allows it to be
treated collinearly.

Every parameter introduced above refers to a factorization scale, the
mixing scheme fixes the shapes at $\mu_{0}=1~\mathrm{GeV}$, and the
LCDAs are needed at the factorization scale of the convolution.  The twist-2 quark
LCDAs are evolved by separating each flavor component into its
octet and singlet projections.  The octet projection renormalizes
multiplicatively in the standard Efremov-Radyushkin-Brodsky-Lepage
manner~\cite{Lepage:1980fj,Efremov:1979qk}, while the singlet one mixes
with $\phi_{g}$ under the coupled two-by-two kernel of
Refs.~\cite{Kroll:2002nt,Ball:2007hb,Agaev:2014wna}, whose leading-order
anomalous dimensions we take from Ref.~\cite{Kroll:2002nt}.  The twist-3
parameters run as well, the chiral masses with the inverse of the
running quark mass and $\eta_{3r}$ and $\omega_{3r}$ with the anomalous
dimensions of Refs.~\cite{Ball:1998je,Ball:2006wn}.  All these quantities are
evolved from $\mu_{0}$ to that scale.  Since only the singlet projection feels the gluon,
the common shape imposed at $\mu_{0}$ does not survive evolution, and
differences between the $\eta$ and the $\etap$ are generated
radiatively, at the 20\% level on the singlet moment between
$1$ and $2.4~\mathrm{GeV}$ for $|a_{2}^{g}|\lesssim10$~\cite{Ball:2007hb}.

\subsection{Hard amplitudes}
\label{subsec:hard}

\begin{figure}[!htb]
  \centering
  \includegraphics[width=0.87\textwidth]{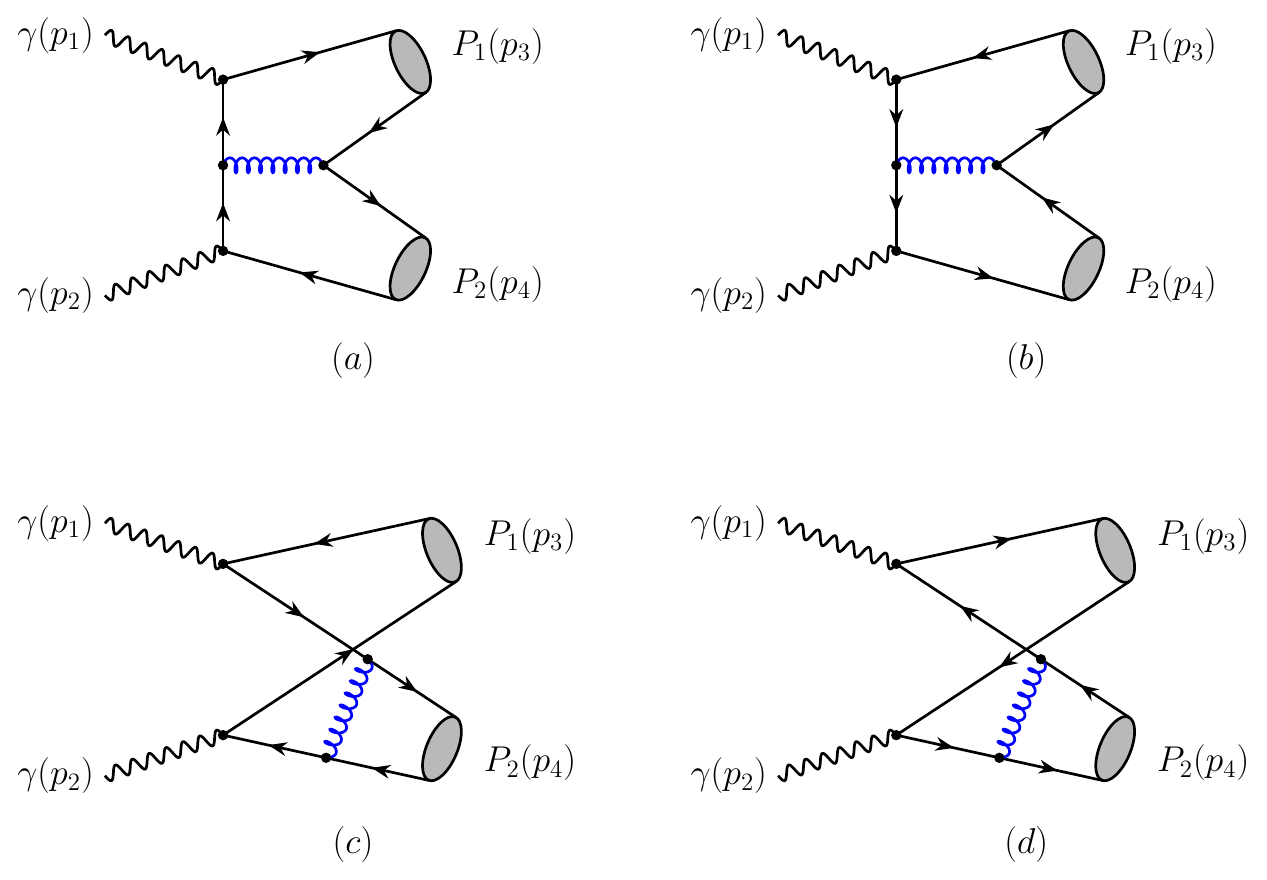}
  \caption{The four typical diagrams for the quark content of
    $\gam(p_{1})\gam(p_{2})\to P_{1}(p_{3})P_{2}(p_{4})$, one for each of
    the groups $a$--$d$, which differ in the virtuality of the internal
    gluon shown in blue.  The gluon attaches to a single quark line in
    $(a)$ and $(b)$ and is exchanged between the two lines in $(c)$ and
    $(d)$.  Each diagram stands
    for six, obtained by moving the gluon attachment and by exchanging
    the two photon legs, $24$ in all.  The topology is common
    to every channel considered here, with $P_{1}P_{2}$ equal to
    $\pieta$, $\pietap$, $\etaeta$, $\etaetap$, or $\etapetap$.}
  \label{fig:feynq}
\end{figure}

At the parton level the quark-content amplitude is built from the
subprocess
$\gam(p_{1})\gam(p_{2})\to q(k_{1})\bar q(k_{2})\,q(k_{3})\bar q(k_{4})$,
in which the two photons couple to the quark lines and a single gluon
binds the two $q\bar q$ pairs.  Each quark line runs through both
mesons, so that a quark flavor contributes only if it is a valence
flavor of $P_{1}$ and of $P_{2}$ alike.  The diagrams fall into four
topological groups $a$, $b$, $c$, and $d$, distinguished by the
virtuality of the internal gluon, which attaches to a single quark line
in groups $a$ and $b$ and is exchanged between the two lines in groups
$c$ and $d$.  The four typical diagrams are shown in
Fig.~\ref{fig:feynq}.  Each of them stands for six, generated by moving
the gluon attachment along the quark lines and by exchanging the two
photon legs, so that $24$ diagrams contribute in total.  The topology is
the same for all five channels, and only the identity of the outgoing
mesons changes.

\begin{figure}[!htb]
  \centering
  \includegraphics[width=0.87\textwidth]{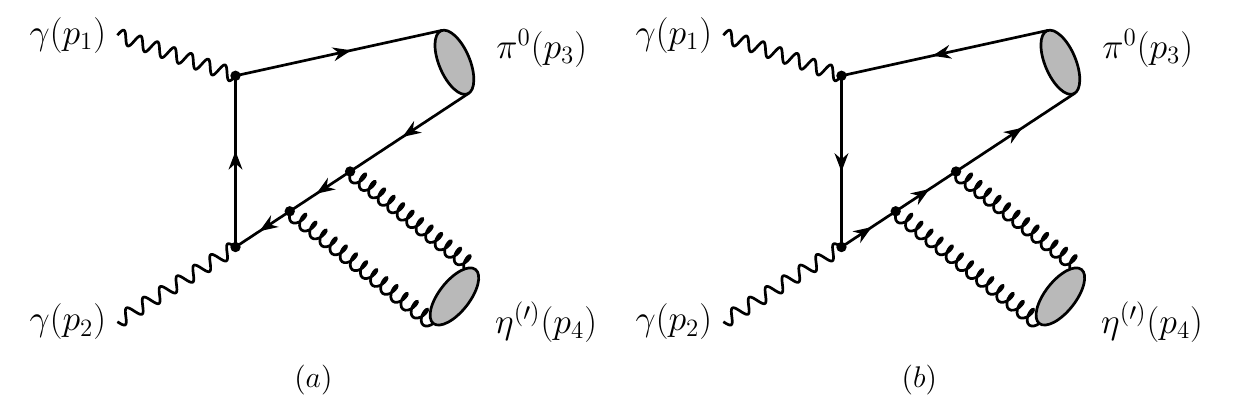}
  \caption{The two typical diagrams for the two-gluon content of
    $\gam(p_{1})\gam(p_{2})\to\pi^{0}(p_{3})\etapp(p_{4})$.  Only the
    $\etapp$ can be gluonic, so the $gg$ pair always forms the lower
    meson.  Each diagram stands for six, obtained from the two gluon
    attachment points and the exchange of the photon legs, giving $12$
    in all.}
  \label{fig:feyng-pi}
\end{figure}

\begin{figure}[!htb]
  \centering
  \includegraphics[width=0.87\textwidth]{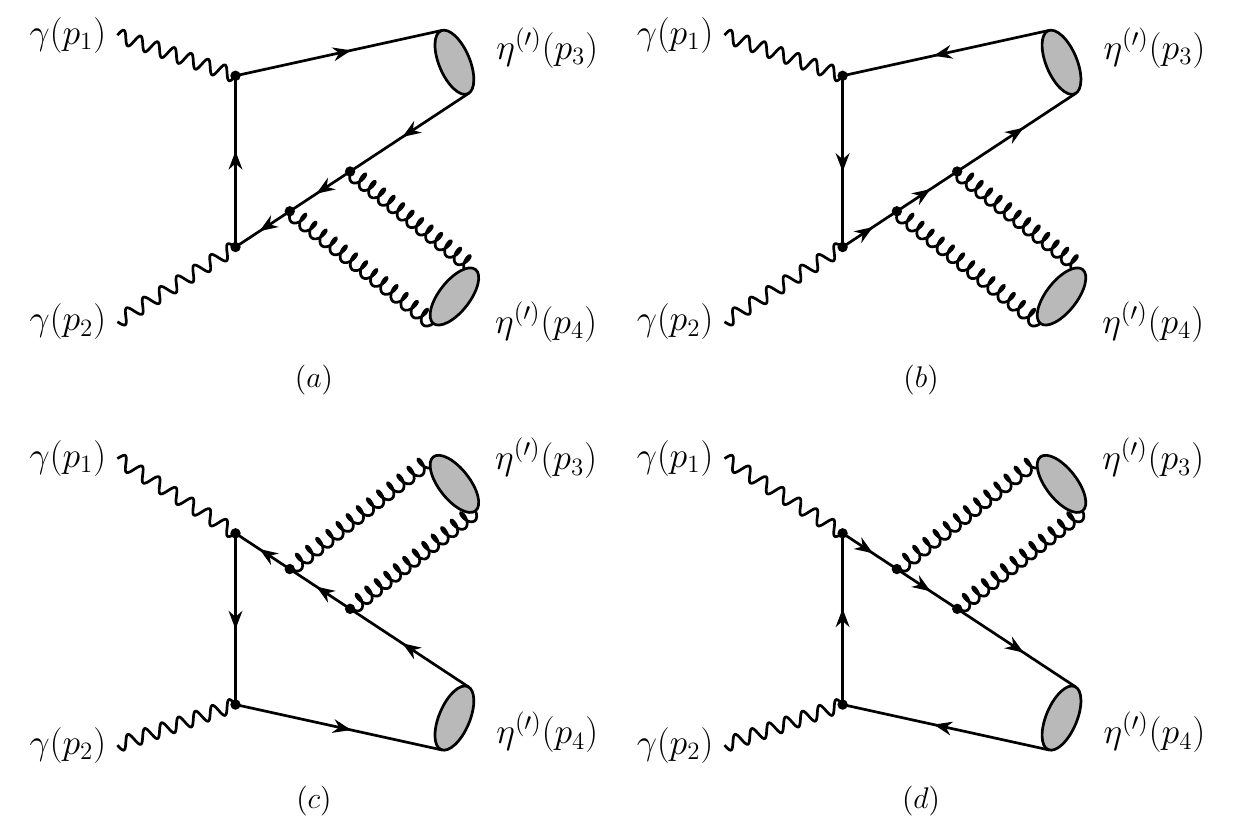}
  \caption{The four typical diagrams for the two-gluon content of
    $\gam(p_{1})\gam(p_{2})\to\etapp(p_{3})\etapp(p_{4})$.  Either meson
    may be gluonic, the $gg$ pair forming the lower one in $(a)$ and
    $(b)$ and the upper one in $(c)$ and $(d)$.  Each diagram again
    stands for six, giving $24$ in all.}
  \label{fig:feyng-eta}
\end{figure}

The two-gluon content of the $\etapp$ contributes through the diagrams
of Figs.~\ref{fig:feyng-pi} and~\ref{fig:feyng-eta}, in which a $gg$
pair radiated from the quark line forms one of the mesons and the other
meson is formed by the quark and antiquark at the two ends of that same
line.  In $\pietapp$ only the $\etapp$ can be gluonic, which leaves two
typical diagrams and $12$ in total, whereas in $\etapp\etapp$ either
meson may be gluonic, giving four typical diagrams and $24$ in total.

The valence partons of $P_{1}$ carry the longitudinal fractions $x$ and
$\xbar=1-x$ and the transverse momentum $\kperpi{1}$, those of $P_{2}$
carry $y$ and $\ybar=1-y$ and the transverse momentum $\kperpi{2}$
\begin{equation}
\label{eq:partonmomenta}
\begin{aligned}
k_{1}^{\mu}&=(2x\omega,\,0,\,\kperpi{1}),
&\quad
k_{2}^{\mu}&=(2\xbar\omega,\,0,\,-\kperpi{1}),\\
k_{3}^{\mu}&=(0,\,2y\omega,\,\kperpi{2}),
&\quad
k_{4}^{\mu}&=(0,\,2\ybar\omega,\,-\kperpi{2}).
\end{aligned}
\end{equation}
Since $t$ and $u$ are fixed by $Q$ and $\theta^{*}$, every ratio of hard
scales appearing below is a function of $x$, $y$, and $\theta^{*}$ alone.

Because the two photons couple to a quark line of definite flavor $q$,
the natural objects are the flavor components of the mesons rather than
the mesons themselves.  We write the coupling of the quark flavor $q$ to
the meson $P$ as $F_{P}^{(q)}$
\begin{equation}
\label{eq:Fq}
F_{\pi^{0}}^{(u)}=-F_{\pi^{0}}^{(d)}=\frac{f_{\pi}}{\sqrt2},
\qquad
F_{\etapp}^{(u)}=F_{\etapp}^{(d)}=\frac{f_{\etapp}^{q}}{\sqrt2},
\qquad
F_{\etapp}^{(s)}=f_{\etapp}^{s},
\end{equation}
with $F_{\pi^{0}}^{(s)}=0$.  Fourier transforming Eqs.~\eqref{eq:pionME}, \eqref{eq:etaME2},
and~\eqref{eq:etaME3} to momentum space and restoring the transverse
momentum gives the spinor projector of the flavor-$q$
component~\cite{Beneke:2000wa,Wei:2002iu,Kroll:2018uvl,Wang:2019trq}
\begin{equation}
\label{eq:projector}
M^{P,q}_{\alpha\beta}=
\frac{iF_{P}^{(q)}}{4}
\left\{\slashed{p}\,\gamma_{5}\Psi_{r}^{\,\pi}
-\mu_{r}\gamma_{5}\!
\left[\Psi_{r}^{p}
-i\sigma_{\mu\nu}\frac{p^{\mu}\bar p^{\nu}}{p\cdot\bar p}
 \frac{\Psi_{r}^{\sigma'}}{6}
+i\sigma_{\mu\nu}p^{\mu}\frac{\Psi_{r}^{\sigma}}{6}
\frac{\partial}{\partial k_{\perp\nu}}\right]\right\}_{\!\alpha\beta},
\end{equation}
where $P=\pi^{0}$, $\eta$, or $\etap$, $\Psi_{r}$ and $\mu_{r}$ are the
wave function and the chiral mass of the flavor state $\pi^{0}$, $\eta_{q}$, or $\eta_{s}$, $\mu_{r}$ being $\mupi$, $\muq$, or $\mus$,
$\bar p^{\mu}=(p^{0},-\mathbf{p})$ and
$\Psi_{r}^{\sigma'}=\partial\Psi_{r}^{\sigma}/\partial x$.  The wave
functions are the LCDAs of Sec.~\ref{subsec:lcda} times a
transverse profile of the Brodsky-Huang-Lepage
form~\cite{Brodsky:1981jv,Lepage:1982gd,Jakob:1993iw,Bolz:1996wh}.  The
convolution is carried out in impact-parameter space, where $b$ is the
transverse separation of the quark and the antiquark inside the meson,
conjugate to $\kperp$, and $1/b$ acts as the factorization scale that
separates the soft wave function from the hard kernel.  There the
wave functions read
\begin{equation}
\label{eq:bspaceSigma}
\widehat{\Psi}_{r}^{\,i}(x,b,\mu)=4\pi\,\phi_{r}^{\,i}(x,\mu)
\exp\!\left[-\frac{x\xbar\,b^{2}}{4\beta_{r}^{2}}\right],
\qquad i=\pi,p,\sigma,
\end{equation}
the oscillator parameter $\beta_{r}$ being fixed by the mean intrinsic
transverse momentum $\langle\kperp^{2}\rangle_{r}$ of the flavor
state~\cite{He:2026bfv}.  The LCDAs are taken at the scale $1/b$ and
evolved as described in Sec.~\ref{subsec:lcda}, so that the twist-2
shapes of the $\etapp$ components acquire a mild dependence on the parent
meson through the singlet evolution.  For the $\eta_{q}$ and $\eta_{s}$ we let
$\langle\kperp^{2}\rangle_{r}$ scale with the square of the decay
constant, $\langle\kperp^{2}\rangle_{r}=\langle\kperp^{2}\rangle_{\pi}
(f_{r}/f_{\pi})^{2}$, which keeps the probability of the valence Fock
state the same for the three flavor states, as in the analyses of the
$\etapp$ transition form factors~\cite{Feldmann:1998vh,Kroll:2002nt}.

With these wave functions the quark-content amplitude follows in $k_{T}$
factorization~\cite{Botts:1989kf,Li:1992nu,Li:1998is,Kurimoto:2001zj,Nagashima:2002ia}
as a sum over the quark flavors
\begin{equation}
\label{eq:Mquark}
\mathcal{M}^{\,q\bar q}_{\lambda_{1}\lambda_{2}}
=\sum_{q=u,d,s}e_{q}^{2}\,F_{P_{1}}^{(q)}F_{P_{2}}^{(q)}\,
 \mathcal{A}^{(q)}_{\lambda_{1}\lambda_{2}},
\end{equation}
in which each flavor contributes with its charge squared and with the
product of its couplings to the two mesons $P_{1}$ and $P_{2}$.  The
selection rule stated
above is built into Eq.~\eqref{eq:Mquark}, a flavor that is absent from
either meson dropping out through a vanishing $F^{(q)}$.  The reduced
amplitude is the convolution of the two wave functions with the hard kernels
\begin{equation}
\label{eq:Aq}
\begin{aligned}
\mathcal{A}^{(q)}_{\lambda_{1}\lambda_{2}}=&
\int_{0}^{1}\!\mathrm{d}x\!\int_{0}^{1}\!\mathrm{d}y\!
   \int_{0}^{\infty}\!\frac{b\,\mathrm{d}b}{(4\pi)^{2}}
   \!\!\sum_{i,j\in\{\pi,p,\sigma,\sigma'\}}\!\!
   \sum_{n\in\{a,b,c,d\}}\!\!
   \widehat{\Psi}_{r_{2}}^{\,i}(y,b,1/b)\,
   \TildeT^{\lambda_{1}\lambda_{2}}_{nij}(x,y,Q,\theta^{*},b,t_{n})\\
&\times\widehat{\Psi}_{r_{1}}^{\,j}(x,b,1/b)\,
   S_{t}(x)\,S_{t}(y)\,\exp[-S(x,y,Q,b,t_{n})],
\end{aligned}
\end{equation}
where $r_{1}$ and $r_{2}$ are the flavor states of $P_{1}$ and $P_{2}$
that contain $q$, and the first label of the kernel refers to the LCDA of
$P_{2}$ and the second to that of $P_{1}$.  The transverse momenta are
kept only in the internal gluon propagator, where they regulate the
end point region, so that the kernels localize at $\bb_{1}+\bb_{2}=0$ and
a single impact parameter
survives~\cite{Li:2010nn,Hu:2012cp,He:2026bfv}.  The strong coupling of
group $n$ is evaluated at the hard scale $t_{n}$, set by the virtuality
of the internal gluon of that group and bounded from below by $1/b$, a
choice that minimizes the higher-order
corrections~\cite{Li:1992nu,Kroll:2018uvl}.  The Sudakov factor $\exp[-S]$
resums the double logarithms from soft and collinear gluon emission and
suppresses the nonperturbative region of large
$b$~\cite{Botts:1989kf,Li:1992nu}, while the jet function $S_{t}$ resums
the threshold logarithms that the twist-3 LCDAs, nonvanishing at
the end points, would otherwise
generate~\cite{Li:1998is,Kurimoto:2001zj,Li:2001ay}.  Both are taken over
unchanged from the pion and kaon channels, and the detailed expressions
of $S$, $S_{t}$, and $t_{n}$ are given in Appendix~\ref{app:hard}.

The kernels $\TildeT^{\lambda_{1}\lambda_{2}}_{nij}$ are the same
functions of $x$, $y$, $\theta^{*}$, and $b$ as in the $\pizpiz$
channel~\cite{He:2026bfv}, the charge and flavor structure having been
moved into Eq.~\eqref{eq:Mquark}.  Twist 2 enters through the prefactor
\begin{equation}
\label{eq:kappa1}
\kappa_{1}=\pi^{2}\alpha_{\mathrm{em}}\,\alpha_{s}(t_{n})
\end{equation}
and twist 3 through
\begin{equation}
\label{eq:kappa2}
\kappa_{2}=\frac{\mu_{r_{1}}\mu_{r_{2}}}{\omega^{2}}\,\kappa_{1},
\end{equation}
which carries the chiral masses of both flavor components, evaluated at
$\mu=1/b$.  As
representative examples we give the kernels of group $a$, in which the
internal gluon has the virtuality $s_{a}=4\xbar y\,\omega^{2}>0$ and
enters through $\mathrm{F}(s_{a},b)=2\pi\Hone(\sqrt{s_{a}}\,b)$.  At
twist 2 they read
\begin{align}
\label{eq:Ta2}
\TildeT^{++}_{a\pi\pi}&=
\frac{4}{9}\kappa_{1}\,
\frac{xy+\xbar\ybar}{s^{2}c^{2}\,x\ybar}\,\mathrm{F}(s_{a},b),
\nonumber\\
\TildeT^{+-}_{a\pi\pi}&=
\frac{4}{9}\kappa_{1}
\left[4+\frac{s^{2}(\xbar y+x\ybar)}{c^{2}x\ybar}
       +\frac{c^{2}(\xbar y+x\ybar)}{s^{2}x\ybar}\right]\mathrm{F}(s_{a},b),
\end{align}
and at twist 3 the pseudoscalar-pseudoscalar kernels read
\begin{align}
\label{eq:Ta3}
\TildeT^{++}_{app}&=
-\frac{2}{9}\kappa_{2}
\left[2-\frac{1-\xbar y}{s^{2}x\ybar}-\frac{1-\xbar y}{c^{2}x\ybar}\right]
\mathrm{F}(s_{a},b),
\nonumber\\
\TildeT^{+-}_{app}&=
\frac{2}{9}\kappa_{2}
\left[\frac{1}{s^{2}c^{2}}-\frac{2(1-x\ybar)}{\xbar y}\right]
\mathrm{F}(s_{a},b).
\end{align}
The full set, with the Bessel-function factors that the
$\partial/\partial k_{\perp\nu}$ term of Eq.~\eqref{eq:projector}
generates in groups $c$ and $d$ and the discrete symmetries relating the
four groups, is given in Appendix~\ref{app:hard}.

Two features of the twist-3 sector shape the results.  The ratio
$\kappa_{2}/\kappa_{1}=4\mu_{r_{1}}\mu_{r_{2}}/Q^{2}$ measures the
strength of the twist-3 amplitude relative to the twist-2 one.  It
contains no quark mass, since each meson supplies one chirality flip and
the two compensate each other along the quark lines, and with chiral
masses set by the quark condensate it is formally a power correction but
numerically of order one or larger for $Q\simeq2$--$4~\mathrm{GeV}$.  The
two convolutions differ in addition at the end points.  The Sudakov and
threshold factors suppress the regions where a parton carries almost no
momentum, but there the twist-3 LCDA $\phi_{r}^{p}$ stays finite
while the twist-2 one vanishes, so an end point contribution survives at
twist 3 alone.  The twist-3 amplitude therefore exceeds the twist-2 one
over much of this range, as in the pion and kaon
channels~\cite{Gorsky:1989ev,Wang:2019trq,He:2026bfv}.  The chiral masses
also fix how the enhancement is shared among the flavor components.  With
$\muq=\mupi$ the nonstrange component of an $\etapp$ carries exactly the
chiral enhancement of the $\pi^{0}$, and with $\mus$ nearly equal to
$\muq$ the strange component carries almost the same, its twist-3 part
differing only in shape through the mass correction $\rho_{s}^{2}$ of
Eq.~\eqref{eq:DAexp}.

The amplitudes above are built from the flavor components of the mesons,
and the physical channels follow from the rotation of
Eq.~\eqref{eq:statemix}.  This rotation is already contained in the
couplings of Eq.~\eqref{eq:Fq}, through the decay constants of
Eq.~\eqref{eq:fmix}, so that Eq.~\eqref{eq:Mquark} applies to the
physical channels as it stands.  The $u$ and the $d$ quark couple to the
same flavor states, so their reduced amplitudes coincide,
$\mathcal{A}^{(u)}_{\lambda_{1}\lambda_{2}}
=\mathcal{A}^{(d)}_{\lambda_{1}\lambda_{2}}$, and with
$F_{\pi^{0}}^{(d)}=-F_{\pi^{0}}^{(u)}$ and $F_{\etapp}^{(d)}=F_{\etapp}^{(u)}$
the flavor sum reduces to
\begin{equation}
\label{eq:Mpieta}
\mathcal{M}^{\,q\bar q}_{\lambda_{1}\lambda_{2}}
=\bigl(e_{u}^{2}-e_{d}^{2}\bigr)\,F_{\pi^{0}}^{(u)}F_{P}^{(u)}\,
 \mathcal{A}^{(u)}_{\lambda_{1}\lambda_{2}}
\end{equation}
for $P_{1}P_{2}=\pi^{0}P$ with $P=\eta,\etap$, and to
\begin{equation}
\label{eq:Metaeta}
\mathcal{M}^{\,q\bar q}_{\lambda_{1}\lambda_{2}}
=\bigl(e_{u}^{2}+e_{d}^{2}\bigr)\,F_{P_{1}}^{(u)}F_{P_{2}}^{(u)}\,
 \mathcal{A}^{(u)}_{\lambda_{1}\lambda_{2}}
+e_{s}^{2}\,F_{P_{1}}^{(s)}F_{P_{2}}^{(s)}\,
 \mathcal{A}^{(s)}_{\lambda_{1}\lambda_{2}}
\end{equation}
for $P_{1},P_{2}=\eta,\etap$.  The relative sign between the two
light-quark couplings of the $\pi^{0}$ is not a convention but the
physical distinction between the isovector $\pi^{0}$ and the isoscalar
$\eta_{q}$.  Both photons couple to the same quark line, so each flavor
enters weighted by $e_{q}^{2}$ and the flavor sum inherits the sign of
the state.  A $\pi^{0}$ in the final state therefore brings
$e_{u}^{2}-e_{d}^{2}=1/3$ where the nonstrange component of an $\etapp$
brings $e_{u}^{2}+e_{d}^{2}=5/9$, a factor $3/5$ before any decay
constant or mixing factor is applied.  The strange component enters
through the second term of Eq.~\eqref{eq:Metaeta} alone.  By
Eqs.~\eqref{eq:Fq} and~\eqref{eq:fmix} its weight relative to the
nonstrange term is $\tfrac25(f_{s}/f_{q})^{2}$ times $\tan^{2}\phi$ in
$\etaeta$, $\cot^{2}\phi$ in $\etapetap$ and $-1$ in $\etaetap$.  Numerically
this amounts to about $0.4$, $0.9$, and $-0.6$, so the small charge $e_{s}^{2}$ is largely compensated by
$f_{s}>f_{q}$.  The strange component is a sizable correction in
$\etaeta$, competes on equal terms in $\etapetap$, where the mixing
favors the $\eta_{s}$, and cancels part of the nonstrange one in
$\etaetap$.

The two-gluon amplitude has a different structure.  The $gg$ pair is
radiated from the quark line itself, so the diagrams of
Figs.~\ref{fig:feyng-pi} and~\ref{fig:feyng-eta} contain no internal
gluon propagator, and the transverse momenta enter only the quark
propagators, where they are subleading.  Together with the quadratic
end point suppression of $\phi_{g}$ this allows the amplitude to be
computed collinearly, with the LCDAs and the coupling taken at
$\mu=Q$~\cite{Kroll:2002nt,Ball:2007hb}.  The quark meson is now formed
by the two ends of a single quark line, along which the vector couplings
conserve chirality, so only its chiral-even twist-2 projection
contributes and the twist-3 amplitudes drop out.  The two-gluon content
feeds $\mathcal{M}_{+-}$ alone, its equal-helicity kernel vanishing
identically, in agreement with earlier
calculations~\cite{Atkinson:1983yh,Harland-Lang:2013ncy}.  For a gluonic
$P_{2}$ we find
\begin{equation}
\label{eq:Mgluon}
\mathcal{M}^{\,gg}_{+-}
=-\frac{2i}{9\sqrt3}\,
 \frac{\pi^{2}\alpha_{\mathrm{em}}\alpha_{s}(Q)}{\omega^{2}}\,
 f_{P_{2}}^{1}(Q)\sum_{q=u,d,s}e_{q}^{2}F_{P_{1}}^{(q)}
 \int_{0}^{1}\!\mathrm{d}x\!\int_{0}^{1}\!\mathrm{d}y\,
 \phi_{r_{1}}^{\,\pi}(x,Q)\,\phi_{g}(y,Q)\,K(x,y),
\end{equation}
with the kernel
\begin{equation}
\label{eq:Kgluon}
K(x,y)=\frac{xy^{2}-\xbar\ybar^{2}-2s^{2}c^{2}(x+y-1)}
 {x\xbar y\ybar\,\bigl(s^{2}xy+c^{2}\xbar\ybar\bigr)
  \bigl(c^{2}xy+s^{2}\xbar\ybar\bigr)}.
\end{equation}
For a gluonic $P_{1}$ the same expression holds with the roles of the two
mesons and of $x$ and $y$ interchanged, and in $\etapp\etapp$ both terms
are present.  Once convoluted with the antisymmetric $\phi_{g}$, which
removes the parts of $K$ even under $y\to\ybar$, Eq.~\eqref{eq:Kgluon}
coincides with the kernels of Refs.~\cite{Atkinson:1983yh,Harland-Lang:2013ncy}
up to constant factors that reflect the different normalizations of the
two-gluon projector and of $\phi_{g}$.  The two factors in its denominator
are positive throughout the physical region, so the quark propagators
never go on shell, the convolution develops no absorptive part and, with
the end point factors supplied by the two LCDAs, needs no regularization.

The quark meson enters Eq.~\eqref{eq:Mgluon} through the same charge-weighted coupling $e_{q}^{2}F_{P_{1}}^{(q)}$ as in Eq.~\eqref{eq:Mquark},
so the flavor selection acts on the gluonic amplitude in the same way,
the $\pi^{0}$ contributing in proportion to $e_{u}^{2}-e_{d}^{2}$.  The
gluonic meson, on the other hand, enters through the singlet coupling of
Eq.~\eqref{eq:f1def}, whose two terms add in the $\etap$ and largely
cancel in the $\eta$, so that the ratio $f_{P}^{1}/f_{P}^{q}$ is about
7 times larger for the $\etap$ than for the $\eta$.  In the $\pi^{0}$
channels, where the quark amplitudes of Eq.~\eqref{eq:Mpieta} stand in the
ratio $F_{\etap}^{(u)}/F_{\eta}^{(u)}=\tan\phi$, the channels $\pietap$ and
$\pieta$ therefore differ, apart from the singlet evolution of the twist-2
LCDAs, only through their two-gluon content, which is far more
visible in $\pietap$.  The two-gluon amplitude is of twist-2 strength
without chiral enhancement and is linear in $a_{2}^{g}$, whose sign is
not known.  Its
effect is therefore an interference in $\mathcal{M}_{+-}$ of
undetermined sign, which is the reason the two-gluon content is treated
as a theoretical uncertainty.

\subsection{Soft handbag contribution and coherent sum}
\label{subsec:soft}

At present energies the meson pair can also be formed through the soft
overlap of the produced quark and antiquark with the final mesons, a
contribution that a consistent factorization at subleading power has to
keep alongside the chirally enhanced twist-3 one~\cite{Kivel:2013fja}.
In the pion and kaon channels the measured energy dependence of the
neutral-to-charged ratios was largely reproduced only once such a contribution
was added coherently to the hard amplitude~\cite{He:2026bfv}.  Unlike the
hard amplitudes, its normalization and phase are not computed but have
to be taken from data.  We describe it by the soft handbag
mechanism~\cite{Diehl:2001fv,Diehl:2009yi}, in which the two photons
couple to a single quark line, and the quark-antiquark pair produced in
the subprocess $\gam\gam\to q\bar q$ turns into the meson pair through a
soft transition, each meson taking almost the whole momentum of the quark
or antiquark that enters it.  The soft transition is described by the
annihilation form factor $R_{P_{1}P_{2}}(Q^{2})$, and the opposite-helicity
amplitude reads~\cite{Diehl:2001fv,He:2026bfv}
\begin{equation}
\label{eq:Asoft}
\mathcal{M}^{\rm soft}_{+-}
=-\frac{16\pi\alpha_{\mathrm{em}}}{\sin^{2}\theta^{*}}\,R_{P_{1}P_{2}}(Q^{2}).
\end{equation}
The equal-helicity amplitude $\mathcal{M}^{\rm soft}_{++}$ vanishes,
because the soft matrix element selects a massless quark and antiquark of
opposite helicities, which the two photons can produce only with opposite
helicities.  The soft amplitude therefore interferes with the
opposite-helicity part of the hard amplitude alone.

Since both photons couple to the same quark, the form factor is a sum
over the quark flavors with the squared charges as weights
\begin{equation}
\label{eq:Rsum}
R_{P_{1}P_{2}}=\sum_{q=u,d,s}e_{q}^{2}\,R^{q}_{P_{1}P_{2}},
\end{equation}
where $R^{q}_{P_{1}P_{2}}$ is the form factor of the quark part of the
energy-momentum tensor of flavor $q$~\cite{Diehl:2001fv}.  These flavor
form factors were related across channels by $SU(3)$ flavor symmetry in
Ref.~\cite{Diehl:2009yi}.  For the octet mesons two independent form
factors remain, a valence one $R^{u}_{2\pi}$, in which the photons couple
to a quark flavor present in the meson pair, and a nonvalence one
$R^{s}_{2\pi}$, in which they couple to an absent flavor, as for the
strange quark in $\pi\pi$.  The $\etapp$ were described there by the
octet state $\eta_{8}$ alone.  The physical $\eta$ and $\etap$, however,
contain the flavor singlet as well, so we redo the analysis in the
flavor basis of Sec.~\ref{subsec:lcda}, with the meson pairs built from
$\pi^{0}$, $\eta_{q}$, and $\eta_{s}$.

The handbag mechanism itself selects the dominant structure.  Since the
quark that absorbs the photons has to transfer almost all of its momentum
to one meson and its partner to the other, it must be a valence quark of
both.  Structures in which this is not the case are suppressed.  They
comprise the nonvalence form factor of the octet analysis, which was
found to fall off with energy faster than the valence one by
nearly a power of $Q^{2}$~\cite{Diehl:2009yi}, and, once the singlet is
included, further OZI-violating structures in which one of the mesons is
attached to the rest of the diagram only through gluons.  We neglect all of them.  This is the limit in which the
neutral-to-charged kaon ratio $R_{K^{0}\bar K^{0}}=\tfrac25R_{K^{+}K^{-}}$
was used in the analysis of the pion and kaon channels~\cite{He:2026bfv},
and it is the limit in which the soft amplitude obeys the same flavor
selection as the hard quark amplitude of Eq.~\eqref{eq:Mquark}.  The
value of the nonvalence form factor extracted in
Ref.~\cite{Diehl:2009yi}, sizable at present energies, was obtained by
attributing the whole cross section to the soft mechanism, whereas the
chirally enhanced hard amplitude takes a comparable share in this energy
range~\cite{He:2026bfv}.  To express this limit we write each meson in
flavor space as $P=\sum_{q}a_{q}^{P}\,q\bar q$, so that
$(a_{u}^{P},a_{d}^{P},a_{s}^{P})$ equals $(1,-1,0)/\sqrt2$ for the
$\pi^{0}$, $(1,1,0)/\sqrt2$ for the $\eta_{q}$ and $(0,0,1)$ for the
$\eta_{s}$.  The flavor form factors then reduce to the overlap of the valence wave functions
\begin{equation}
\label{eq:Rrule}
R^{q}_{P_{1}P_{2}}=2R^{u}_{2\pi}\,a_{q}^{P_{1}}a_{q}^{P_{2}},
\end{equation}
the source quark entering one meson and its antiquark the other.
Equation~\eqref{eq:Rrule} reproduces the octet relations of Ref.~\cite{Diehl:2009yi} at vanishing nonvalence form factor.  The
derivation and the general expressions are given in
Appendix~\ref{app:soft}.

Combining Eqs.~\eqref{eq:Rsum} and~\eqref{eq:Rrule}, the form factors
of the flavor-state pairs follow in units of
$R_{\pizpiz}=\tfrac59R^{u}_{2\pi}$,
\begin{equation}
\label{eq:Rflavor}
R_{\pi^{0}\eta_{q}}=\tfrac35R_{\pizpiz},
\qquad
R_{\eta_{q}\eta_{q}}=R_{\pizpiz},
\qquad
R_{\eta_{s}\eta_{s}}=\tfrac25R_{\pizpiz},
\qquad
R_{\pi^{0}\eta_{s}}=R_{\eta_{q}\eta_{s}}=0.
\end{equation}
Each of these has a simple origin.  The pair $\eta_{q}\eta_{q}$ has the
valence content of $\pizpiz$, the relative sign of $u\bar u$ and $d\bar d$
in the $\pi^{0}$ dropping out of the product, and the two form factors
are equal.  In $\pi^{0}\eta_{q}$ that sign survives, the $u$ and $d$
quarks entering with $e_{u}^{2}-e_{d}^{2}$ instead of
$e_{u}^{2}+e_{d}^{2}$, which gives the same factor $3/5$ as in the hard
amplitude.  The $\eta_{s}$ is a pure $s\bar s$ state, so $\eta_{s}\eta_{s}$
carries the small charge $e_{s}^{2}$ but without the dilution by
$(1/\sqrt2)^{2}$ that each flavor suffers in $\pizpiz$, hence $2/5$
rather than $1/5$.  The pairs $\pi^{0}\eta_{s}$ and $\eta_{q}\eta_{s}$
share no valence flavor and vanish, exactly as their hard quark
amplitudes do.

For the physical channels the rotation of Eq.~\eqref{eq:statemix} is
applied to the form factors of the flavor-state pairs, as it was to the
hard amplitude in Sec.~\ref{subsec:hard}.  With $R_{\pi^{0}\eta_{s}}$ and
$R_{\eta_{q}\eta_{s}}$ vanishing, Eq.~\eqref{eq:Rflavor} makes the form
factors of all five channels proportional to that of $\pizpiz$
\begin{equation}
\label{eq:lambda}
R_{P_{1}P_{2}}=\lambda_{P_{1}P_{2}}\,R_{\pizpiz},
\end{equation}
with the flavor factors
\begin{equation}
\label{eq:lambdavalues}
\begin{aligned}
\lambda_{\pieta}&=\tfrac35\cos\phi,
&\qquad
\lambda_{\etaeta}&=\cos^{2}\phi+\tfrac25\sin^{2}\phi,
&\qquad
\lambda_{\etaetap}&=\tfrac35\sin\phi\cos\phi,\\
\lambda_{\pietap}&=\tfrac35\sin\phi,
&\qquad
\lambda_{\etapetap}&=\sin^{2}\phi+\tfrac25\cos^{2}\phi.
\end{aligned}
\end{equation}
These are the charge and flavor weights of Eqs.~\eqref{eq:Mpieta}
and~\eqref{eq:Metaeta}, normalized to $\pizpiz$, in the limit
$f_{q}=f_{s}=f_{\pi}$ with all reduced amplitudes equal, in which the
soft and the hard amplitudes share one flavor structure.  The
soft factors depend on the mixing angle alone, whereas the hard
amplitudes depart from them through $f_{q}$, $f_{s}$, the different
dynamics of the strange component and the two-gluon term, so the balance
between the two mechanisms changes from channel to channel.  Both are
built with the same rotation, so the signs of the flavor factors carry
the relative sign of the soft and the hard contributions from one
channel to the next.  The soft and the two-gluon amplitudes leave the
equal-helicity amplitude to the hard quark content,
$\mathcal{M}_{++}=\mathcal{M}^{\,q\bar q}_{++}$, and all three
mechanisms meet in the opposite-helicity amplitude
\begin{equation}
\label{eq:Mpmsum}
\mathcal{M}_{+-}=\mathcal{M}^{\,q\bar q}_{+-}+\mathcal{M}^{\,gg}_{+-}
+\mathcal{M}^{\rm soft}_{+-},
\end{equation}
where they interfere.  The hard amplitudes carry phases generated by
the timelike internal propagators of the convolution, while the cross
sections fix the phase of the soft amplitude only relative to them, so
no separate soft-to-hard phase is introduced.  We write the soft
amplitude of Eq.~\eqref{eq:Asoft} as
\begin{equation}
\label{eq:Msoft}
\mathcal{M}^{\rm soft}_{+-}
=\frac{16\pi\alpha_{\mathrm{em}}}{\sin^{2}\theta^{*}}\,
 \lambda_{P_{1}P_{2}}\,|R_{\pizpiz}(Q^{2})|\,
 e^{i\delta_{P_{1}P_{2}}(Q)},
\end{equation}
with the modulus fixed by Eq.~\eqref{eq:lambda} and the sign of
Eq.~\eqref{eq:Asoft} included in the phase.  Besides the phase of the
annihilation form factor and its phase relative to the hard amplitudes,
$\delta_{P_{1}P_{2}}$ contains the phase acquired when the $q\bar q$ pair
reaches the final state through $s$-channel resonances.  Which
resonances contribute depends on the isospin and flavor content of the
pair, so the phase differs from channel to channel even where flavor
symmetry relates the moduli, as found for the pion and kaon
channels~\cite{He:2026bfv}.

\section{Numerical analysis}
\label{sec:numerics}

\subsection{Input parameters}
\label{subsec:inputs}

The perturbative inputs, the strong coupling with its thresholds and the
resummation factors, are those of the pion and kaon
channels~\cite{He:2026bfv} and are given in Appendix~\ref{app:hard}.
The meson masses are taken from Ref.~\cite{ParticleDataGroup:2024cfk}
and enter only through the chiral masses and the quark-mass parameters
of the twist-3 LCDAs, and the pion decay constant is
$f_{\pi}=130.2~\mathrm{MeV}$~\cite{ParticleDataGroup:2024cfk}.

For the mixing parameters we use the lattice determination of the
Extended Twisted Mass Collaboration at the physical
point~\cite{Ottnad:2025zxq}
\begin{equation}
\label{eq:mixinputs}
f_{q}=138.6\pm4.4~\mathrm{MeV},
\qquad
f_{s}=170.7\pm3.3~\mathrm{MeV},
\qquad
\phi=39.3^{\circ}\pm2.0^{\circ}.
\end{equation}
The angle coincides with the phenomenological value of the
Feldmann-Kroll-Stech analysis~\cite{Feldmann:1998vh}.  The nonstrange coupling lies
about 6\% above $f_{\pi}$, as the OZI rule leads one to expect for
a state with the quark content of the pion, while $f_{s}/f_{q}=1.23$
measures the $SU(3)$ breaking of the strange component.  By
Eqs.~\eqref{eq:fmix} and~\eqref{eq:f1def} these values give
$f_{\eta}^{q}=107~\mathrm{MeV}$ and $f_{\etap}^{q}=88~\mathrm{MeV}$ for
the quark couplings of the $\pi^{0}$ channels, and $f_{\eta}^{1}=25~\mathrm{MeV}$
and $f_{\etap}^{1}=148~\mathrm{MeV}$ for the singlet couplings, the near
cancellation in $f_{\eta}^{1}$ being the reason why the two-gluon
content shows up mainly in the $\etap$ channels.  The three parameters
are varied independently within their errors, and the soft flavor
factors of Eq.~\eqref{eq:lambdavalues} follow the variation of $\phi$.

The twist-2 shapes are taken $SU(3)$ symmetric at
$\mu_{0}=1~\mathrm{GeV}$, with the same coefficients for the $\pi^{0}$, the $\eta_{q}$, and the $\eta_{s}$
\begin{equation}
\label{eq:a2a4inputs}
a_{2}^{r}=0.20\pm0.07,
\qquad
a_{4}^{r}=0.10\pm0.10,
\end{equation}
the values of the pion and kaon channels~\cite{He:2026bfv}, which cover
the spread between the lattice result~\cite{Bali:2019dqc} and the
light cone sum rule
determinations~\cite{Khodjamirian:2011ub,Khodjamirian:2017fxg,Cheng:2020vwr}.
The fits to the $\etapp$ transition form factors in
Ref.~\cite{Agaev:2014wna} find $a_{2}=0.10$--$0.25$ for the two flavor
states, within the same range, so the data give no indication of $SU(3)$
breaking in the twist-2 shapes beyond what evolution generates.  Since
the twist-3 amplitudes dominate in the present energy range, these
coefficients affect the cross sections less than they do in the
transition form factors.

The two-gluon coefficient is the least constrained input.  The
next-to-leading-order fit to the transition form factors gives
$a_{2}^{g}(\mu_{0})=9\pm12$~\cite{Kroll:2002nt}, a combined analysis with
the inclusive decay $\Upsilon(1S)\to\etap X$ gives $4.6\pm2.5$ at
$\mu^{2}=2~\mathrm{GeV}^{2}$~\cite{Ali:2003kq}, and the light cone
sum rule analysis of the transition form factors finds $-7.5$ to
$-9.3$~\cite{Agaev:2014wna}.  Even its sign is thus not established, and
we adopt the conservative range
\begin{equation}
\label{eq:a2ginput}
a_{2}^{g}(2.4~\mathrm{GeV})=0\pm20
\end{equation}
used in Ref.~\cite{Ball:2007hb}, which covers the three determinations above.

The chiral masses follow from Eqs.~\eqref{eq:mupi} and~\eqref{eq:muqs}.
For the light quarks we use the ratio $m_{s}/m_{ud}=27.227\pm0.081$
together with $m_{s}(2~\mathrm{GeV})=93.46\pm0.58~\mathrm{MeV}$, since the
ratio is known far more precisely than the individual light-quark
masses~\cite{FlavourLatticeAveragingGroupFLAG:2024oxs}.  This gives
\begin{equation}
\label{eq:muinputs}
\mupi(2~\mathrm{GeV})=\muq(2~\mathrm{GeV})=2.654\pm0.018~\mathrm{GeV},
\qquad
\mus(2~\mathrm{GeV})=2.510\pm0.016~\mathrm{GeV}.
\end{equation}
The uncertainties are those of the quark masses alone, as appropriate
for quantities fixed by Eqs.~\eqref{eq:mupi} and~\eqref{eq:muqs}.
Since $\muq$ equals $\mupi$ by Eq.~\eqref{eq:muqs} and $\mus$ lies only
slightly below, the three chiral masses are nearly equal, which supports
the approximation of Ref.~\cite{Huang:2026hvs}, where all three are set
equal.  The three-parton parameters of the pion at
$2~\mathrm{GeV}$ are~\cite{Ball:1998je,Ball:2006wn}
\begin{equation}
\label{eq:tw3inputs}
\eta_{3\pi}=0.0127\pm0.004,
\qquad
\omega_{3\pi}=-1.1\pm0.5.
\end{equation}
For the two flavor states we take the three-parton coupling of the
pion, $f_{3q}=f_{3s}=f_{3\pi}$.  The quark masses enter this coupling
only through small corrections of order $m_{q}$, which arise from its
mixing with $f_{P}m_{q}$ under renormalization.  The sum rules confirm
this for the strange quark, giving $f_{3K}=f_{3\pi}$ within
errors~\cite{Ball:2006wn}.  The ratio
$\eta_{3r}=f_{3r}/(f_{r}\mu_{r})$ then follows from the decay constant
and the chiral mass of each state.  For the $\eta_{q}$, with
$\muq=\mupi$, it gives $\eta_{3q}=\eta_{3\pi}f_{\pi}/f_{q}$, which
differs from $\eta_{3\pi}$ by 6\%, far below its uncertainty,
so the $\eta_{q}$ takes over the twist-3 LCDAs of the pion with
$\eta_{3q}=\eta_{3\pi}$.  For the $\eta_{s}$ the larger $f_{s}\mus$ lowers
the ratio by about 20\%, and we keep
$\eta_{3s}=\eta_{3\pi}f_{\pi}\mupi/(f_{s}\mus)=0.010$.  The gluon
momentum-fraction parameter is taken over for both states,
$\omega_{3q}=\omega_{3s}=\omega_{3\pi}$.  The quark masses in Eq.~\eqref{eq:rho} are
the averages of Ref.~\cite{FlavourLatticeAveragingGroupFLAG:2024oxs},
which give $\rho_{s}^{2}=0.074$ at $2~\mathrm{GeV}$.

The transverse profile of the pion wave function is fixed by
$\langle\kperp^{2}\rangle_{\pi}^{1/2}=0.35~\mathrm{GeV}$, the value
required by the $\pi^{0}\to\gam\gam$
constraint~\cite{Brodsky:1981jv,Lepage:1982gd,Wang:2019trq}.  The
scaling with the decay constant of Sec.~\ref{subsec:hard} then gives
$0.37~\mathrm{GeV}$ for the $\eta_{q}$ and $0.46~\mathrm{GeV}$ for the
$\eta_{s}$.  The strange component is thus more compact in impact
parameter space, which makes its convolution less sensitive to the
large-$b$ region where the perturbative treatment is least reliable.

The soft amplitude of Eq.~\eqref{eq:Msoft} is normalized to the
$\pizpiz$ form factor of Ref.~\cite{He:2026bfv}.  Its modulus falls as
$1/Q^{2}$, the behavior expected from dimensional counting and found in
the original handbag analysis~\cite{Diehl:2001fv}.  The constant was
determined in Ref.~\cite{He:2026bfv} by fitting the coherent sum to the
measured ratios of the neutral to the charged pion and kaon cross
sections~\cite{Nakazawa:2004gu,Chen:2006gy,Uehara:2009cka}, with a common
modulus for the pion and kaon form factors, as $SU(3)$ symmetry
implies~\cite{Diehl:2001fv},
\begin{equation}
\label{eq:Rmod}
Q^{2}\,|R_{\pizpiz}|=(0.36\pm0.11)~\mathrm{GeV}^{2}.
\end{equation}
This is about half of the value obtained when the whole cross section
is attributed to the soft mechanism~\cite{Diehl:2001fv}, the remainder
being supplied by the chirally enhanced hard amplitude.  For the phase
we take the linear form
\begin{equation}
\label{eq:deltaQ}
\delta_{P_{1}P_{2}}(Q)=\delta_{0}+\delta_{1}\,(Q-3~\mathrm{GeV}),
\end{equation}
the simplest one that accommodates the rotation of the phase with
energy required in the pion and kaon channels~\cite{He:2026bfv}.  For
$\pizpiz$ we keep $\delta_{0}=246^{\circ}$ and
$\delta_{1}=54^{\circ}/\mathrm{GeV}$, fixed there by the ratio of the
$\pizpiz$ to the $\pi^{+}\pi^{-}$ cross
section~\cite{Uehara:2008ep,Uehara:2009cka} at the modulus of
Eq.~\eqref{eq:Rmod}.  The isovector pairs $\pieta$ and $\pietap$ share
one set of parameters and the isoscalar pairs $\etaeta$, $\etaetap$, and $\etapetap$ another.  They are fitted to the measured ratios of the
$\pieta$ and the $\etaeta$ to the $\pizpiz$ cross
section~\cite{Uehara:2009cf,Uehara:2010mq} and carried over to the
unmeasured channels of each group.

\subsection{Hard contributions}
\label{subsec:hardresults}

With these inputs we first consider the hard mechanism alone.  Its size
at present energies is governed by the two features of the twist-3 sector
identified in Sec.~\ref{subsec:hard}.  With the chiral masses of
Eq.~\eqref{eq:muinputs} the factor $\kappa_{2}/\kappa_{1}$ falls only
from about $7$ to about $2$ between $2$ and $4~\mathrm{GeV}$, and the
end point contribution that survives at twist 3 alone adds to it.

\begin{figure}[!htb]
\centering
\includegraphics[width=0.49\textwidth]{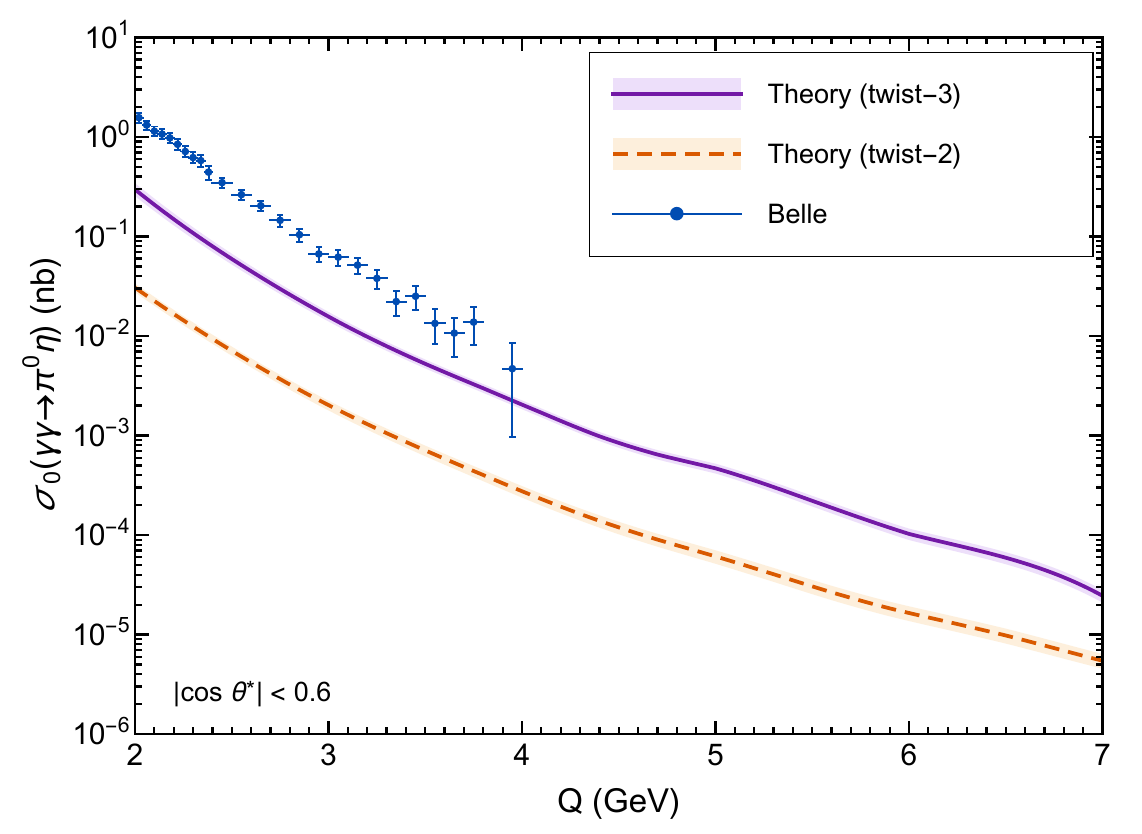}\hfill
\includegraphics[width=0.49\textwidth]{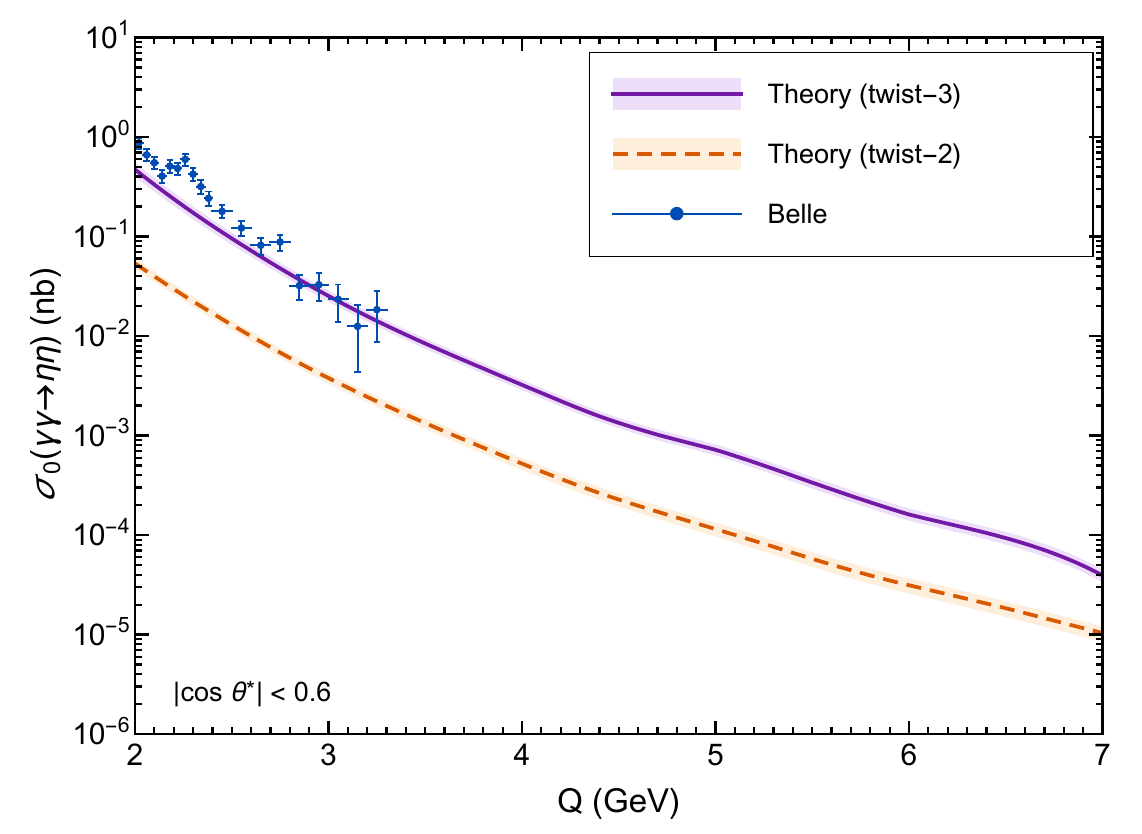}
\caption{Partial cross sections $\sigma_{0}$ ($|\cos\theta^{*}|<0.6$) of
$\gam\gam\to\pieta$ (left) and $\gam\gam\to\etaeta$ (right) from the hard
mechanism, at twist 2 (dashed) and with the two-parton twist-3
amplitudes added (solid).  The twist-2 curves include the two-gluon
amplitude, and the bands show the uncertainty from the hadronic inputs.
The Belle data~\cite{Uehara:2009cf,Uehara:2010mq} are summed over the
same angular range.}
\label{fig:sigmameas}
\end{figure}

The resulting partial cross sections $\sigma_{0}$, integrated over
$|\cos\theta^{*}|<0.6$, of the two measured channels $\pieta$ and
$\etaeta$ are shown as functions of $Q$ in Fig.~\ref{fig:sigmameas}.  The
dashed curves contain the twist-2 quark and two-gluon amplitudes, the
solid ones add the two-parton twist-3 amplitudes, and the bands give the
combined uncertainty from the hadronic inputs of Sec.~\ref{subsec:inputs},
namely the mixing parameters and the parameters of the twist-2 and
twist-3 LCDAs, each varied within its error.  The
Belle points are obtained by summing the published differential cross
sections over the same angular
range~\cite{Uehara:2009cf,Uehara:2010mq}.  The twist-2 cross sections
lie 1 to 2 orders of magnitude below the data, and the twist-3
amplitudes raise them by close to an order of magnitude.  The effective power law
$\sigma_{0}\sim Q^{-n}$ steepens only from $n\simeq6.8$ at twist 2 to
$n\simeq7.2$ at twist 3, so the enhancement is not a threshold effect but
persists over the whole range.  The twist-3 sector thus dominates the hard
cross sections at present energies and brings them to within a factor of
a few of the data, as found for the pion and kaon
channels~\cite{Wang:2019trq,He:2026bfv}.  The two-gluon amplitude, by contrast,
enters through the small singlet coupling $f_{\eta}^{1}$, and varying
$a_{2}^{g}$ over the range of Eq.~\eqref{eq:a2ginput} changes these cross
sections by only about 2\%.

The comparison with the data brings out a difference between the two
channels.  For $\etaeta$ the twist-3 result comes close to the measured
cross section in both size and energy dependence.  It reaches the data
above $2.8~\mathrm{GeV}$ and lies below them by less than a factor of 2
at lower $Q$, and its index $n=7.31\pm0.09$ agrees with the measured
$n=7.8\pm0.6\pm0.4$ for $2.4<Q<3.3~\mathrm{GeV}$~\cite{Uehara:2010mq}.
For $\pieta$ the twist-3 result lies below the measurement by a factor of
4 to 5 between $2.5$ and $3.2~\mathrm{GeV}$ and by a factor of
about 3 above $3.5~\mathrm{GeV}$, and the data fall faster, the Belle
index $n=10.5\pm1.2\pm0.5$ for $3.1<Q<4.1~\mathrm{GeV}$~\cite{Uehara:2009cf}
exceeding our $n=7.07\pm0.02$ in the same range.  Over
$2<Q<4.1~\mathrm{GeV}$ the twist-3 indices of all five
channels cluster around $7.2$, as collected in Table~\ref{tab:nvalues}.
The Belle indices refer to $|\cos\theta^{*}|<0.8$, whereas the twist-3
index depends little on the angular range, since its normalized angular
distribution, discussed below, hardly changes with $Q$.  Deficits of a similar size were found for
$\pizpiz$ and $\KsKs$, and in $\KsKs$ the measured energy dependence is
likewise steeper than the hard one~\cite{He:2026bfv}.

\begin{table}[!htb]
\caption{Effective power-law indices $n$, defined by
$\sigma_{0}\sim Q^{-n}$ and fitted over $2<Q<4.1~\mathrm{GeV}$ at
$|\cos\theta^{*}|<0.6$, for the hard cross sections at twist 2 and
including twist 3.  The errors follow from fitting the edges of the
uncertainty bands.  The Belle indices refer to $|\cos\theta^{*}|<0.8$ and
to $3.1<Q<4.1~\mathrm{GeV}$ for $\pieta$ and $2.4<Q<3.3~\mathrm{GeV}$ for
$\etaeta$, and their two errors are statistical and systematic.}
\label{tab:nvalues}
\begin{ruledtabular}
\begin{tabular}{lccc}
Channel & $n^{\mathrm{tw2}}$ & $n^{\mathrm{tw3}}$ & $n^{\mathrm{Belle}}$\\
\hline
$\pieta$    & $6.82\pm0.06$ & $7.19\pm0.07$ & $10.5\pm1.2\pm0.5$~\cite{Uehara:2009cf}\\
$\pietap$   & $6.93\pm0.05$ & $7.20\pm0.06$ & \\
$\etaeta$   & $6.71\pm0.03$ & $7.20\pm0.07$ & $7.8\pm0.6\pm0.4$~\cite{Uehara:2010mq}\\
$\etaetap$  & $7.39\pm0.25$ & $6.90\pm0.06$ & \\
$\etapetap$ & $6.84\pm0.36$ & $7.25\pm0.18$ & \\
\end{tabular}
\end{ruledtabular}
\end{table}

\begin{figure}[!htb]
\centering
\includegraphics[width=0.329\textwidth]{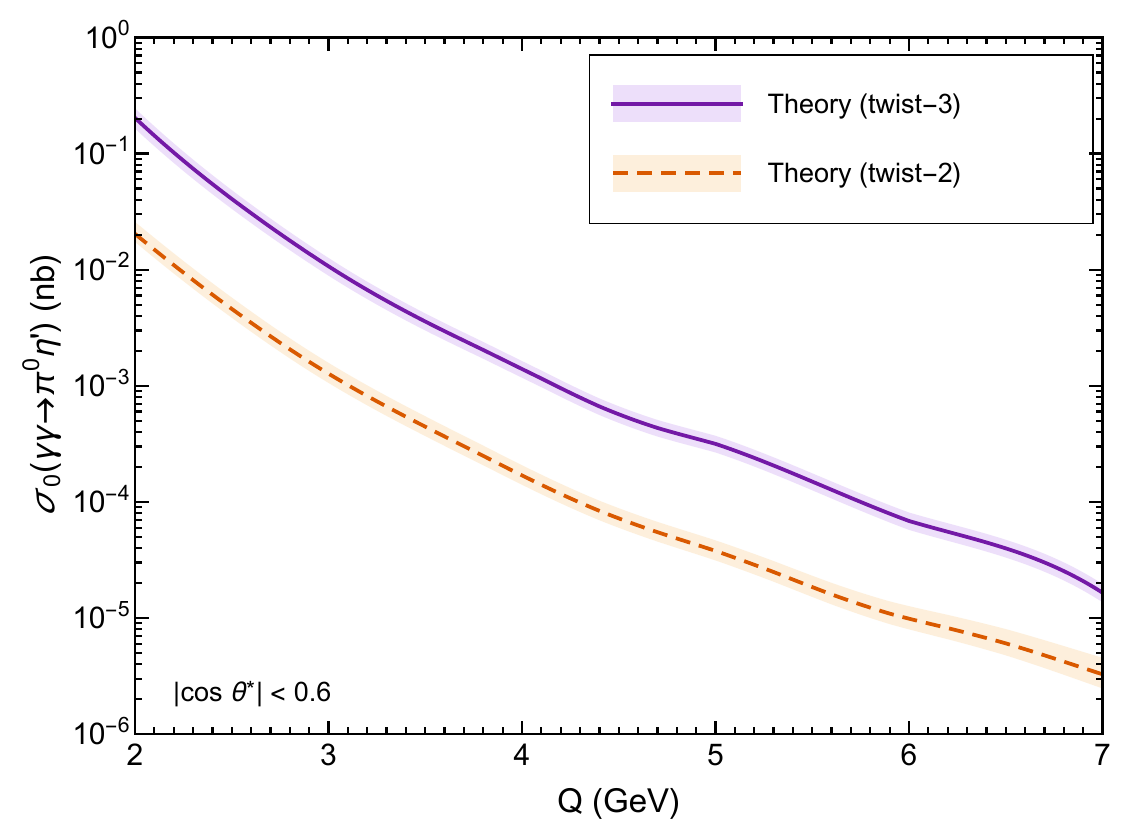}\hfill
\includegraphics[width=0.329\textwidth]{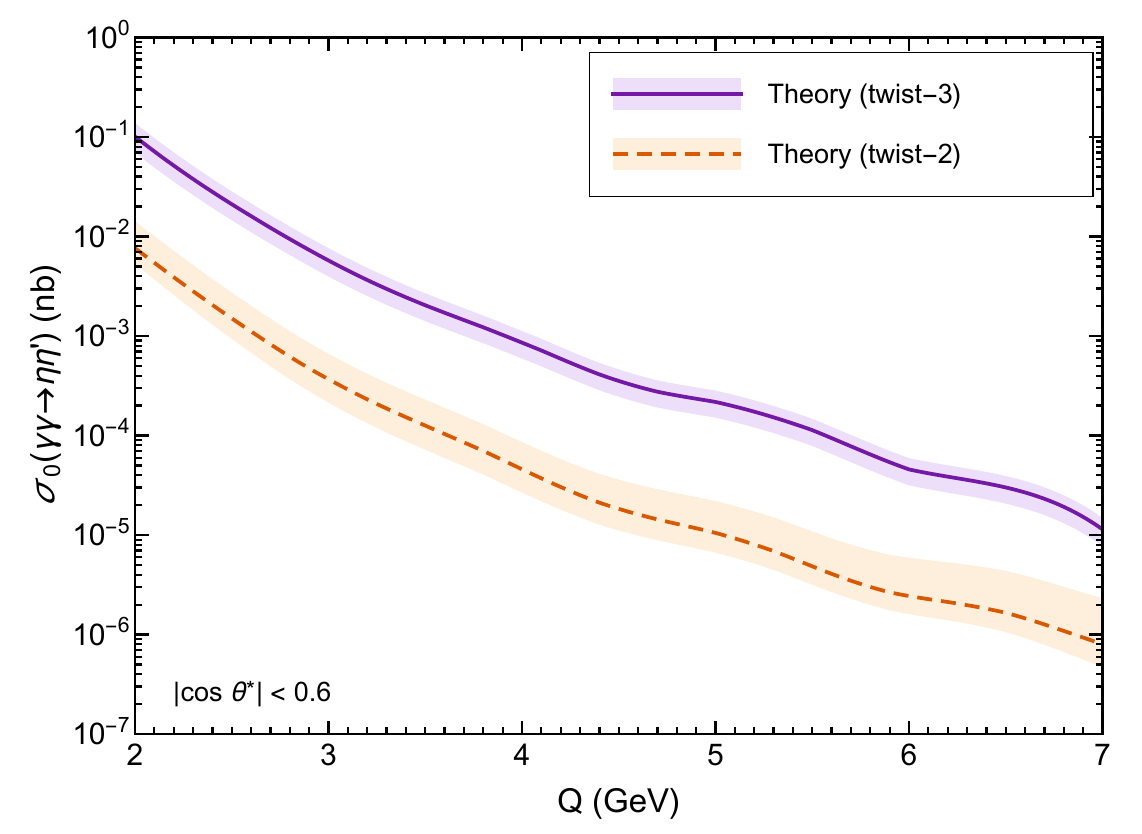}\hfill
\includegraphics[width=0.329\textwidth]{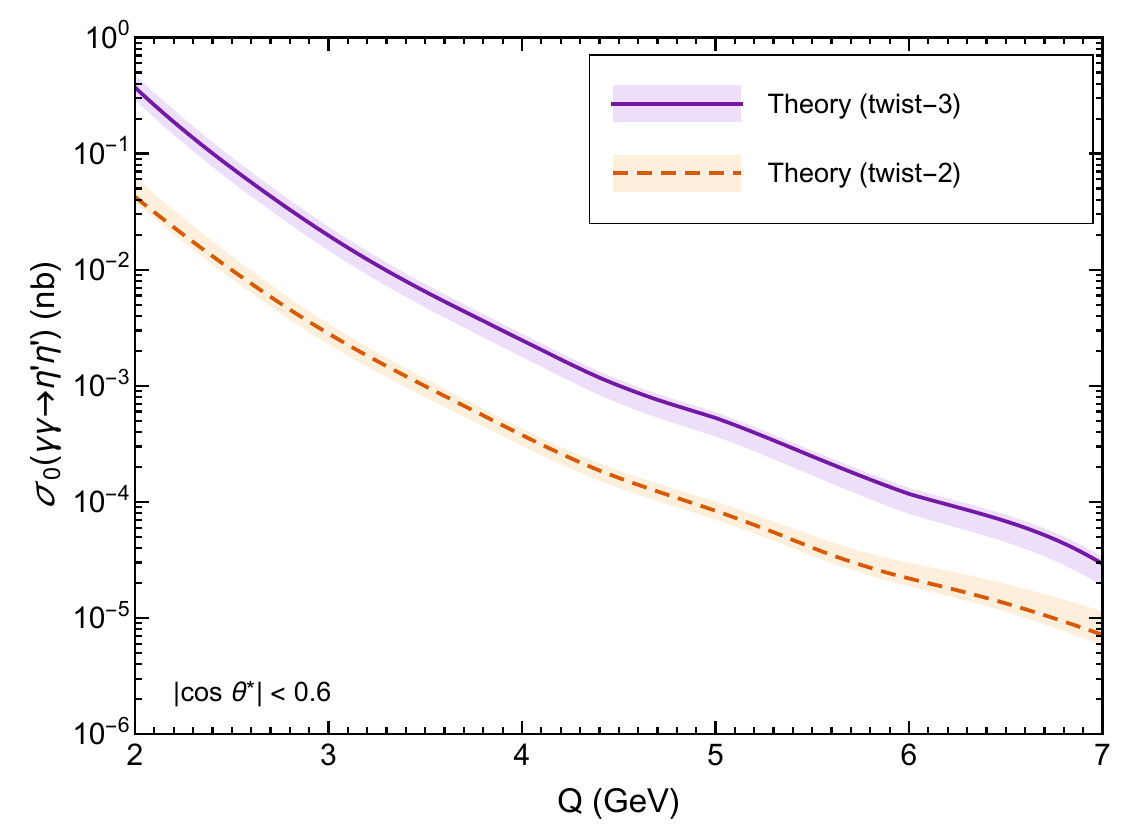}
\caption{Partial cross sections $\sigma_{0}$ ($|\cos\theta^{*}|<0.6$) of
the unmeasured channels $\gam\gam\to\pietap$ (left), $\etaetap$ (middle),
and $\etapetap$ (right) from the hard mechanism.  Curves and bands as in
Fig.~\ref{fig:sigmameas}.}
\label{fig:sigmapred}
\end{figure}

The three unmeasured channels are shown in Fig.~\ref{fig:sigmapred} in
the same way.  Their relative sizes follow the flavor structure of
Sec.~\ref{subsec:hard}.  At $Q=3.05~\mathrm{GeV}$ the twist-3 cross
section of $\pietap$ is $0.68$ times that of $\pieta$, close to
$\tan^{2}\phi=0.67$, as expected from the twist-3 amplitudes of the two
channels being identical up to the ratio
$F_{\etap}^{(u)}/F_{\eta}^{(u)}=\tan\phi$.  The twist-2 ratio,
$0.63$, departs from it through the singlet evolution of the twist-2
LCDAs.  The $\etapetap$ cross section is comparable to the $\etaeta$
one despite the small charge of the strange quark, because the $\etap$ is
mostly $\eta_{s}$ and the strange term of Eq.~\eqref{eq:Metaeta} then
weighs almost as much as the nonstrange one.  The $\etaetap$ channel is
the smallest, at about a quarter of $\etaeta$, because the strange term
enters with the opposite sign and cancels more than half of the
nonstrange one.  What survives the cancellation is sensitive to any
difference between the two terms.  At twist 3 the LCDAs of
Eq.~\eqref{eq:DAexp} belong to the flavor states and do not depend on the
parent meson, and the degree of cancellation hardly changes with $Q$.  At
twist 2 two further differences enter.  The singlet components of the
$\eta$ and the $\etap$ evolve differently from the octet ones, and the
strange and nonstrange wave functions differ in transverse size.  Both
effects grow with $Q$ and deepen the cancellation.  At $3~\mathrm{GeV}$
the twist-2 cross section of $\etaetap$ is only a tenth of the $\etaeta$
one, against a quarter once twist 3 is included.  The twist-3 enhancement
therefore exceeds an order of magnitude in this channel, and $\etaetap$ is the only channel in
Table~\ref{tab:nvalues} whose twist-2 index exceeds the twist-3 one.

The uncertainty bands of these channels are wider and carry the two-gluon
content.  The same variation of $a_{2}^{g}$
changes the twist-3 cross sections at $3~\mathrm{GeV}$ by 13\%, 20\%, and 24\% in $\pietap$, $\etaetap$, and $\etapetap$, against 2\% in the
channels of Fig.~\ref{fig:sigmameas}, as the larger
singlet coupling $f_{\etap}^{1}$ leads one to expect.  In $\etaetap$ the
partial cancellation of the quark amplitude enhances the sensitivity
further.  At twist 2 the gluonic variation reaches nearly 80\% there,
which is why the dashed band is visibly wider, and the same cancellation
makes the channel the most sensitive to the mixing parameters, the
variation of $f_{q}$ alone changing its twist-3 cross section by
25\%.  The $\etap$ channels are
thus where a measurement would constrain the two-gluon content of the
$\etapp$.

\begin{figure}[!htb]
\centering
\includegraphics[width=0.49\textwidth]{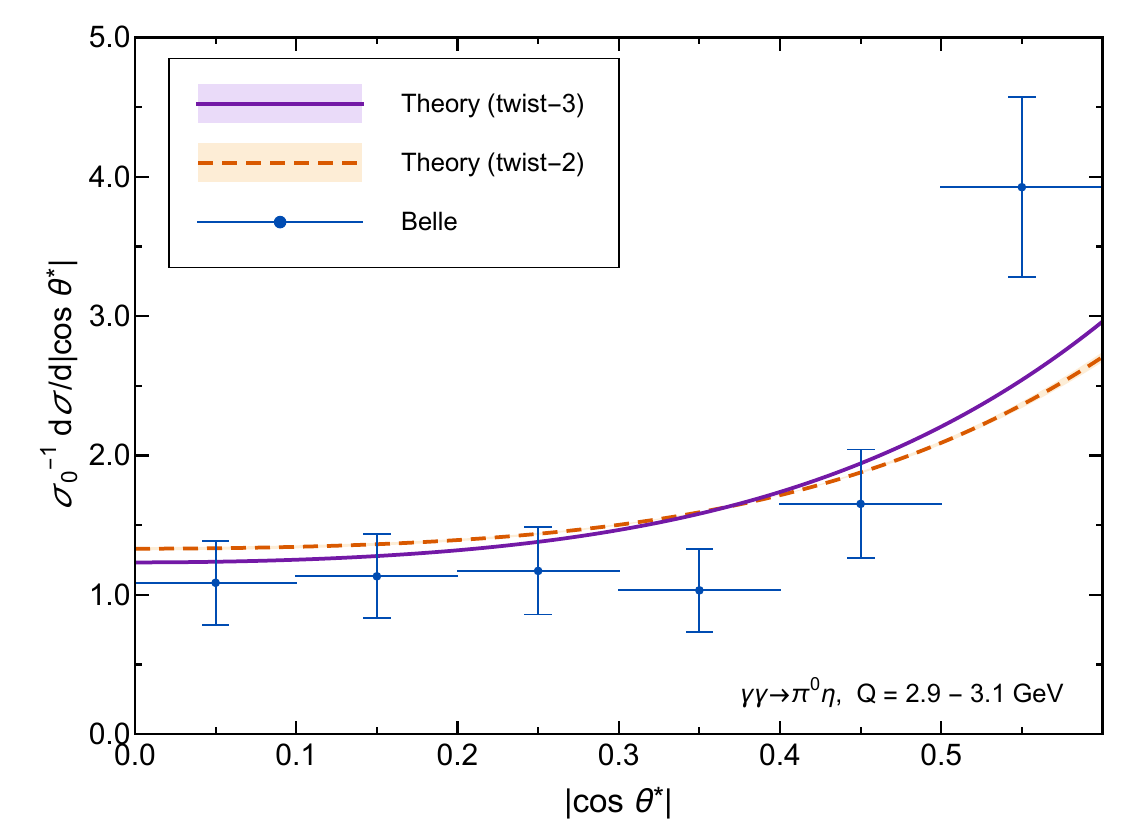}\hfill
\includegraphics[width=0.49\textwidth]{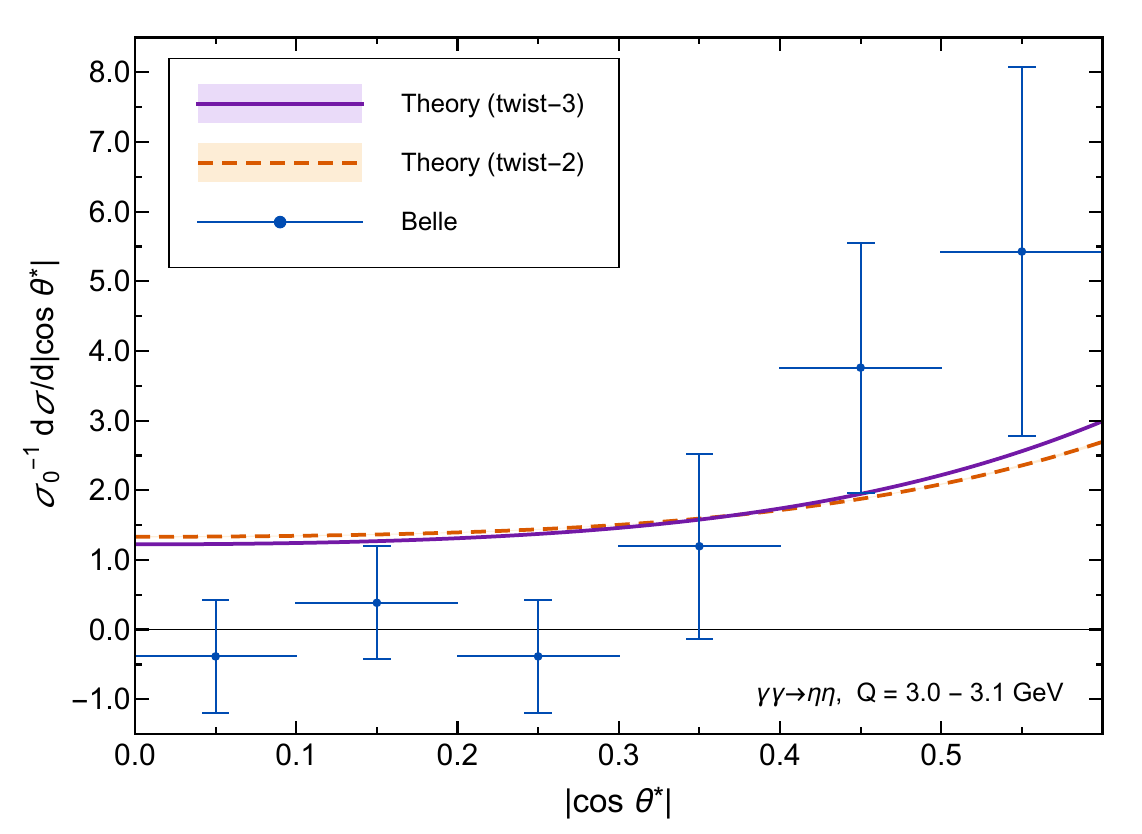}
\caption{Normalized differential cross sections
$\sigma_{0}^{-1}d\sigma/d|\cos\theta^{*}|$ of $\gam\gam\to\pieta$ at
$2.9<Q<3.1~\mathrm{GeV}$ (left) and $\gam\gam\to\etaeta$ at
$3.0<Q<3.1~\mathrm{GeV}$ (right) from the hard mechanism, at twist 2
(dashed) and including twist 3 (solid), compared with the Belle
data~\cite{Uehara:2009cf,Uehara:2010mq}.  The bands from the hadronic
inputs are about 1\% wide and barely visible.}
\label{fig:dsigmanorm}
\end{figure}

The angular dependence is examined in Fig.~\ref{fig:dsigmanorm}, which
shows the normalized distributions
$\sigma_{0}^{-1}d\sigma/d|\cos\theta^{*}|$ in the Belle bins
$2.9<Q<3.1~\mathrm{GeV}$ for $\pieta$ and $3.0<Q<3.1~\mathrm{GeV}$ for
$\etaeta$.  The normalization removes the overall size of the cross
section, and with it most of the theoretical uncertainty.  The input
variations that move $\sigma_{0}$ by 11\% to 13\% in these
channels change the normalized distributions by about 1\%, so the
curves isolate the angular content of the hard kernels.  At twist 3 this content is very close to $1/\sin^{4}\theta^{*}$.
Normalized in the same way, that function takes the values $1.23$ and
$2.69$ at $|\cos\theta^{*}|=0.03$ and $0.57$, against $1.21$ and $2.72$
for the twist-3 curves, while the twist-2 distribution is flatter.  For
$\pieta$ and $\etaeta$ the shape changes by less than 2\% between
$2$ and $4~\mathrm{GeV}$, and the three unmeasured channels differ from
it by a few percent at most.
The $\pieta$ data agree with it except for the last bin, which lies
about two standard deviations higher, in line with the
$\sin^{-4}\theta^{*}$ behavior reported by Belle above
$2.7~\mathrm{GeV}$~\cite{Uehara:2009cf}.  The $\etaeta$ data rise more
steeply, the lower bins being compatible with zero, and Belle finds a
$\sin^{-6}\theta^{*}$ form to describe them better above
$3~\mathrm{GeV}$~\cite{Uehara:2010mq}.  Given the errors this difference
is not yet conclusive.

\subsection{Coherent soft and hard contributions}
\label{subsec:softhard}

\begin{figure}[!htb]
\centering
\includegraphics[width=0.49\textwidth]{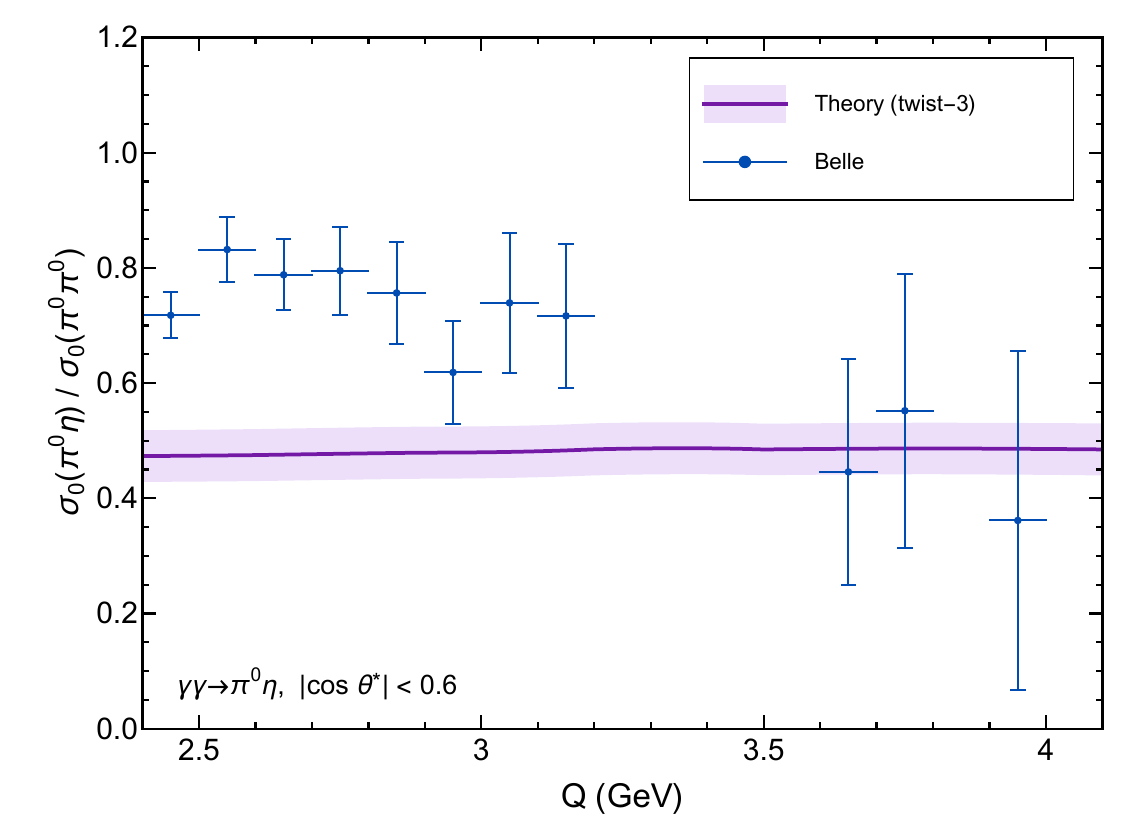}\hfill
\includegraphics[width=0.49\textwidth]{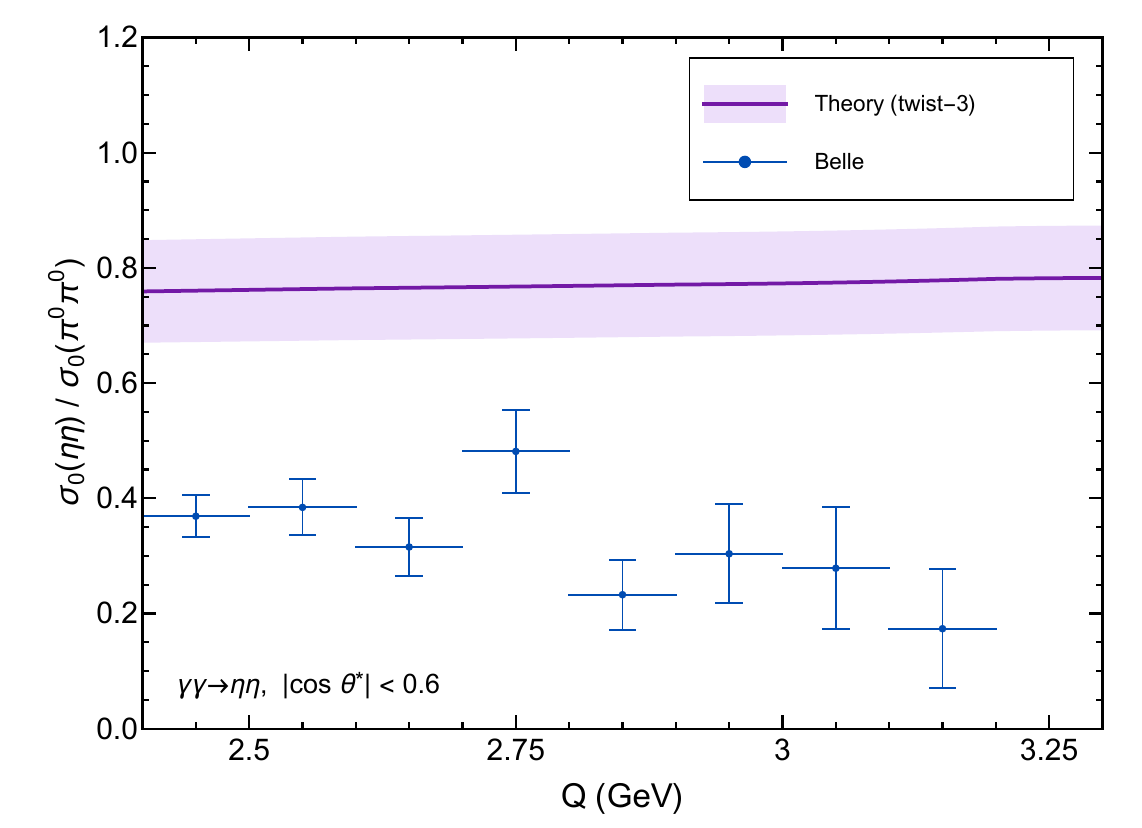}
\caption{Ratios of the partial cross sections $\sigma_{0}$
($|\cos\theta^{*}|<0.6$) of $\gam\gam\to\pieta$ (left) and
$\gam\gam\to\etaeta$ (right) to that of $\gam\gam\to\pizpiz$ from the hard
mechanism including twist 3, compared with the Belle
data~\cite{Uehara:2009cka,Uehara:2009cf,Uehara:2010mq}.  The bands show
the uncertainty from the hadronic inputs.}
\label{fig:ratiohard}
\end{figure}

Belle also quotes the $\pieta$ and $\etaeta$ cross sections relative to
$\pizpiz$~\cite{Uehara:2009cf,Uehara:2010mq}.  In such ratios the
overall size of the hard amplitudes cancels, together with much of the
uncertainty of the inputs the channels share, and what remains is fixed
by the flavor structure of Sec.~\ref{subsec:hard} and by the relative
dynamics of the channels.  The ratios of the hard cross sections
including twist 3 are compared in Fig.~\ref{fig:ratiohard} with the data,
for which the Belle differential cross sections of both channels are summed
over $|\cos\theta^{*}|<0.6$~\cite{Uehara:2009cka,Uehara:2009cf,Uehara:2010mq}.

The hard ratios hardly change with energy and stay near $0.48$ for
$\pieta$ and $0.77$ for $\etaeta$.  The soft
mechanism alone also gives constant ratios.  With the flavor factors of
Eq.~\eqref{eq:lambdavalues} and the statistical factor of
Eq.~\eqref{eq:statfac} they are $2\lambda_{\pieta}^{2}=0.43$ and
$\lambda_{\etaeta}^{2}=0.58$.  Each mechanism falls with energy in the
same way in all channels, so its ratios are fixed by the flavor
structure alone.  The data behave differently.  The measured $\pieta$ ratio is
about $0.75$ below $3.2~\mathrm{GeV}$ and falls to about $0.45$ toward
$4~\mathrm{GeV}$, while the $\etaeta$ ratio is about $0.37$ below
$2.8~\mathrm{GeV}$ and lower beyond.  Below $3.2~\mathrm{GeV}$ the $\pieta$ data thus lie
above the values of both mechanisms and the $\etaeta$ data below them.  Neither
mechanism alone nor an incoherent sum of the two can produce this.  The
soft and hard opposite-helicity amplitudes have to interfere, and the
change of the ratios with energy then points to a relative phase
that varies with $Q$.

We therefore evaluate the ratios with the coherent sum of
Eq.~\eqref{eq:Mpmsum}, taking the soft amplitude of Eq.~\eqref{eq:Msoft}
with the modulus of Eq.~\eqref{eq:Rmod}, and fit the phase of
Eq.~\eqref{eq:deltaQ} to the measured ratios, with the $\pizpiz$ phase
kept at the values of Sec.~\ref{subsec:inputs}.  The fit to the 11
$\pieta$ points gives $\delta_{0}=(220\pm4)^{\circ}$ and
$\delta_{1}=(84\pm11)^{\circ}/\mathrm{GeV}$ for the isovector pairs.
The fit to the 8 $\etaeta$ points gives $\delta_{0}=(279\pm5)^{\circ}$
and $\delta_{1}=(48\pm14)^{\circ}/\mathrm{GeV}$ for the isoscalar
pairs.  Both phases rotate with
energy at a rate of the same size as the pion and kaon phases, $54$ and
$99^{\circ}/\mathrm{GeV}$~\cite{He:2026bfv}.

\begin{figure}[!htb]
\centering
\includegraphics[width=0.49\textwidth]{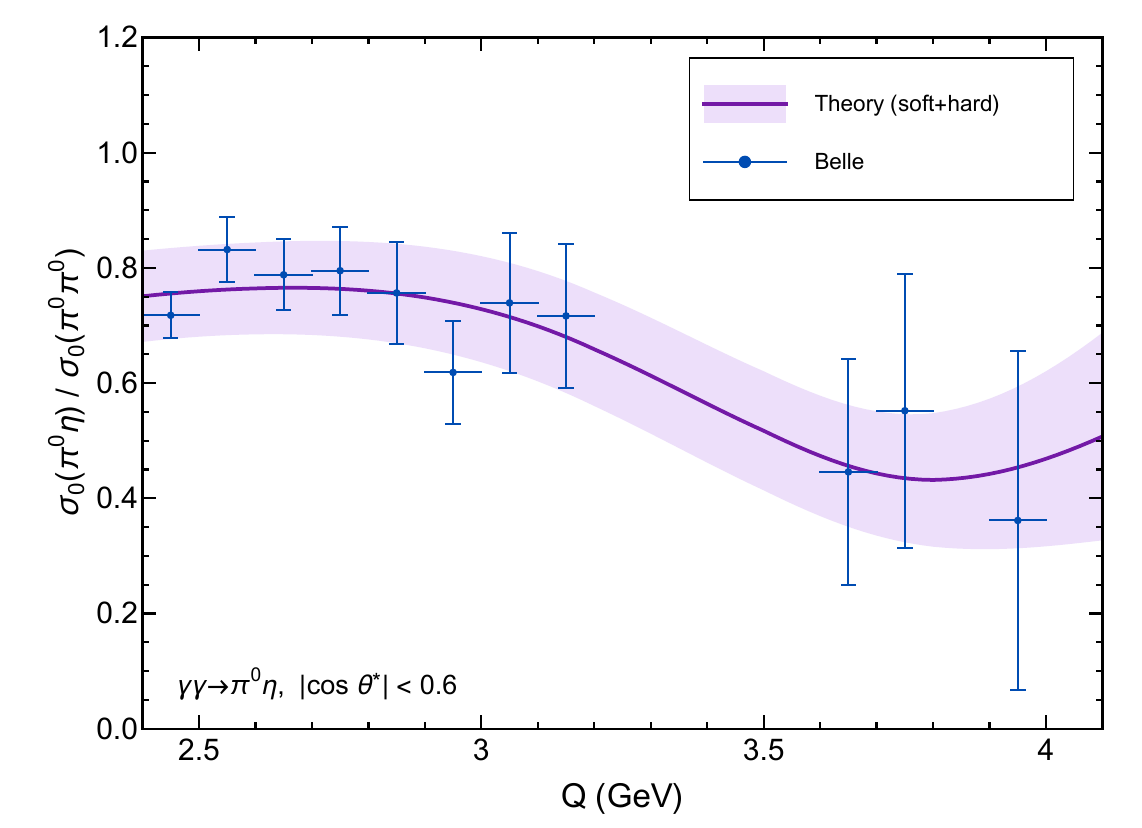}\hfill
\includegraphics[width=0.49\textwidth]{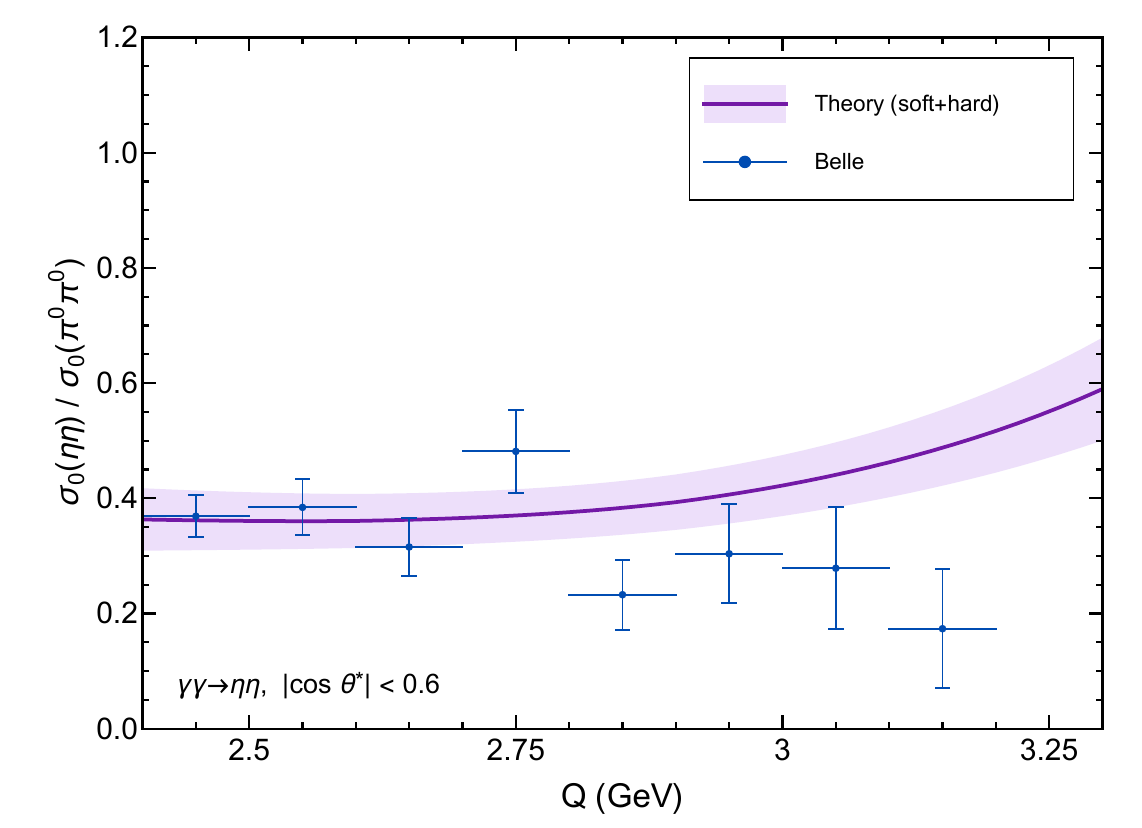}
\caption{Ratios of the partial cross sections $\sigma_{0}$
($|\cos\theta^{*}|<0.6$) of $\gam\gam\to\pieta$ (left) and
$\gam\gam\to\etaeta$ (right) to that of $\gam\gam\to\pizpiz$ for the
coherent sum of the soft and hard amplitudes, compared with the Belle
data~\cite{Uehara:2009cka,Uehara:2009cf,Uehara:2010mq}.  The soft phases
are fitted to these data.  The bands combine in quadrature the
uncertainties of the hadronic inputs of the hard amplitudes, the fitted
soft phase, and the mixing angle in the soft flavor
factors.}
\label{fig:ratiosofthard}
\end{figure}

The ratios obtained with these phases are shown in
Fig.~\ref{fig:ratiosofthard}.  In $\pieta$ the interference is constructive below
$3~\mathrm{GeV}$ and lifts the ratio well above the hard value $0.48$, to
about $0.75$.  As the phase rotates the interference turns destructive,
and near $3.8~\mathrm{GeV}$ the ratio falls to about $0.43$, following
the data over the whole range.  In $\etaeta$ the interference is
destructive and lowers the ratio from the hard value $0.77$ to about
$0.36$ below $2.8~\mathrm{GeV}$, in agreement with the data.  Above
$2.8~\mathrm{GeV}$ the measured ratio falls below the coherent result.
The highest points approach the $\chi_{c0}$ resonance, whose interference
with the continuum may already affect them.  In both channels the energy
dependence of the ratios can thus be understood as the interference of two
mechanisms of comparable size whose relative phase rotates with energy,
as in the pion and kaon channels~\cite{He:2026bfv}.

\begin{figure}[!htb]
\centering
\includegraphics[width=0.329\textwidth]{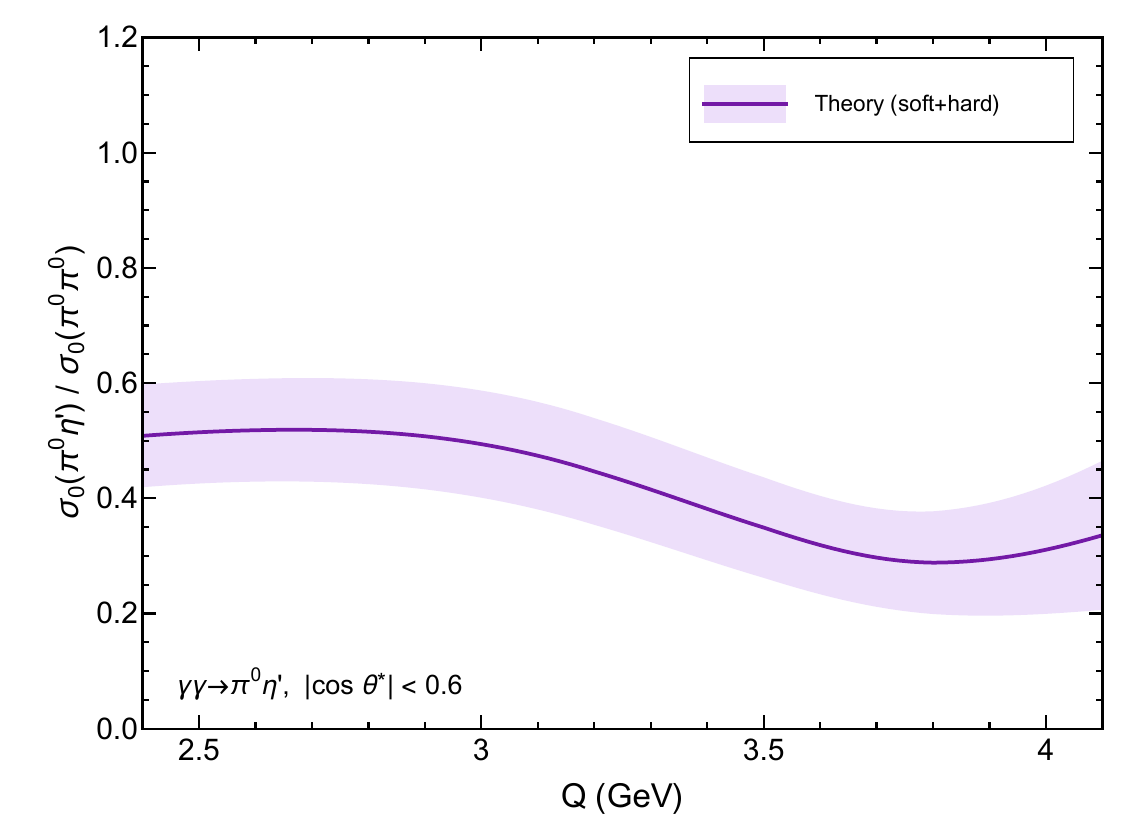}\hfill
\includegraphics[width=0.329\textwidth]{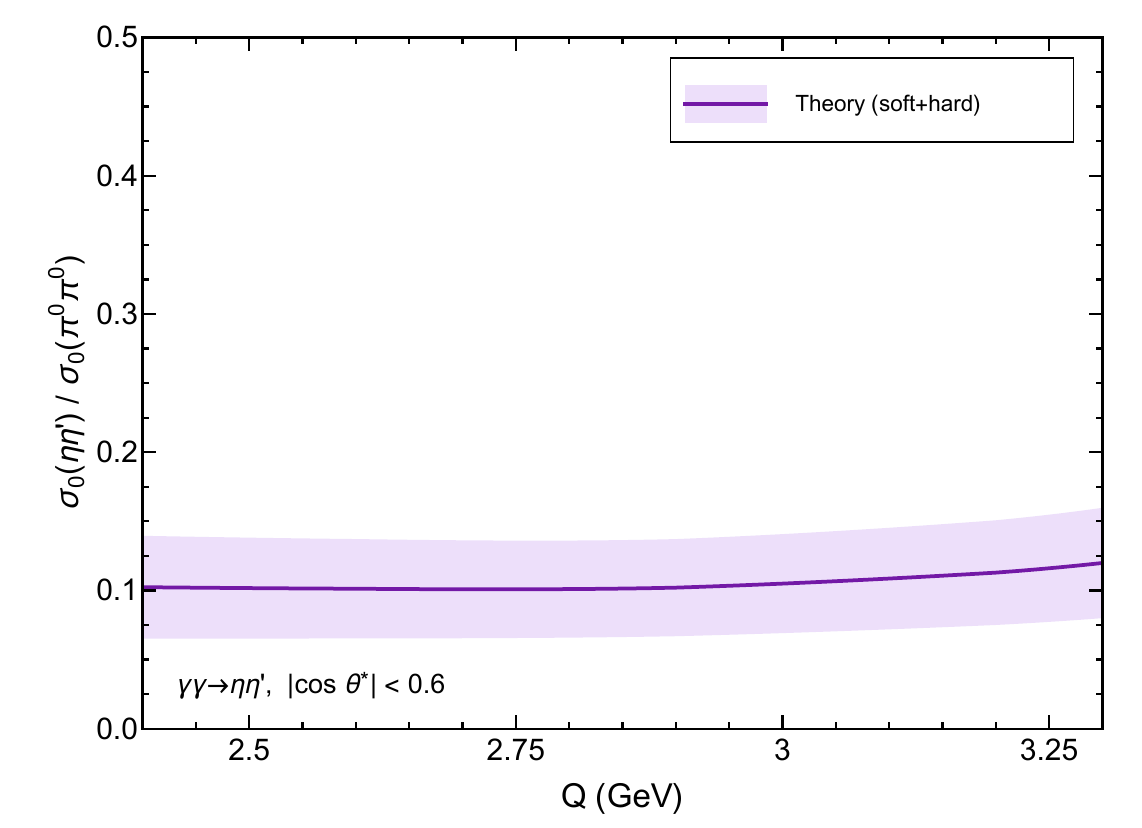}\hfill
\includegraphics[width=0.329\textwidth]{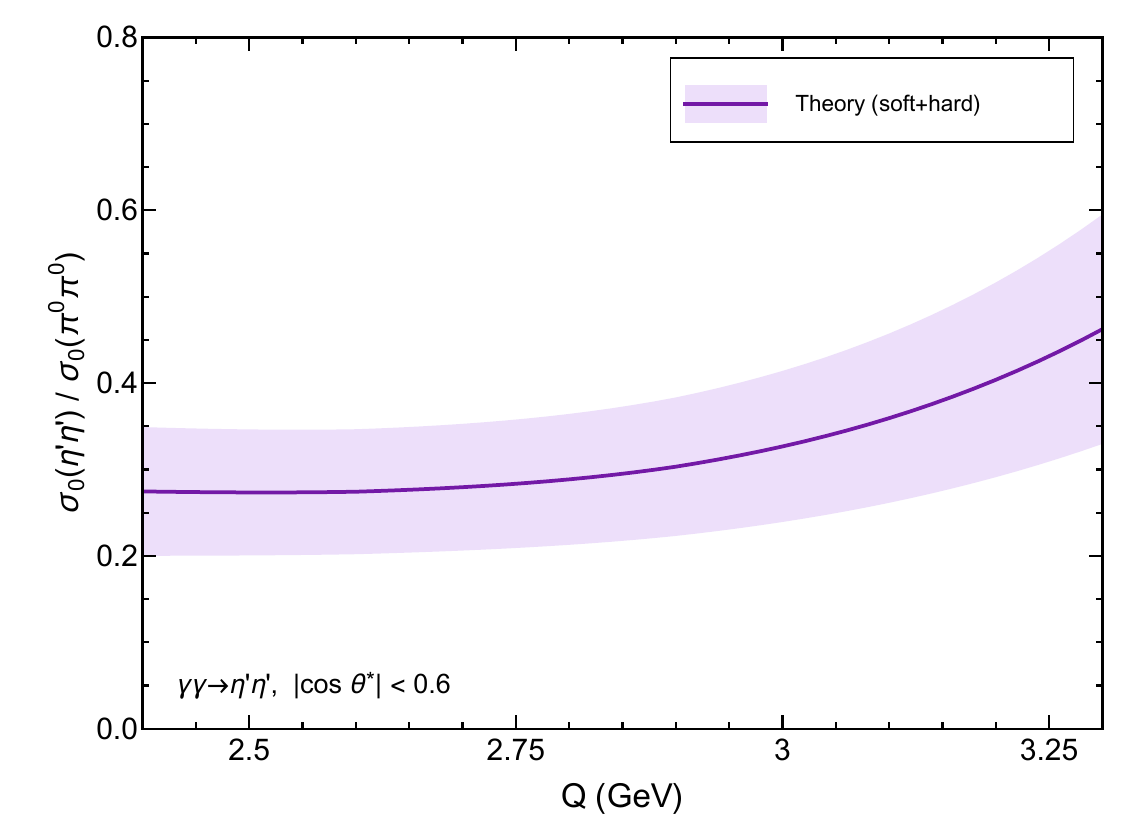}
\caption{Predicted ratios of the partial cross sections $\sigma_{0}$
($|\cos\theta^{*}|<0.6$) of $\gam\gam\to\pietap$ (left), $\etaetap$
(middle), and $\etapetap$ (right) to that of $\gam\gam\to\pizpiz$ for the
coherent sum of the soft and hard amplitudes.  The soft phase of
$\pietap$ is taken from the fit to $\pieta$ and that of $\etaetap$ and
$\etapetap$ from the fit to $\etaeta$.  Bands as in
Fig.~\ref{fig:ratiosofthard}.}
\label{fig:ratiopred}
\end{figure}

For the unmeasured channels the phase of each group is carried over, the
isovector one to $\pietap$ and the isoscalar one to $\etaetap$ and
$\etapetap$.  The resulting ratios are shown in Fig.~\ref{fig:ratiopred}.
The $\pietap$ ratio follows the $\pieta$ one, smaller by a factor close
to $\tan^{2}\phi=0.67$.  The soft
amplitudes of the two channels stand exactly in the ratio $\tan\phi$, and
the hard ones depart from it only through the two-gluon content and the
singlet evolution, so a measurement of $\pietap$ would test the mixing
scheme nearly independently of the soft parameters.  In the isoscalar
channels the interference is destructive as in $\etaeta$.  It lowers the
$\etaetap$ ratio from the hard value $0.17$ to about $0.10$ and the
$\etapetap$ ratio from $0.60$ to about $0.27$ at $2.4~\mathrm{GeV}$, a
suppression that weakens with energy as in $\etaeta$.  The wide band of $\etaetap$ reflects the
partial cancellation of its quark amplitude discussed in
Sec.~\ref{subsec:hardresults}.

With the phases fixed by the ratios, the absolute cross sections follow
without further parameters, and the interference changes their energy
dependence.  Between $3.1$ and $4.1~\mathrm{GeV}$ the coherent $\pieta$
cross section falls with $n=9.7\pm1.6$, against $7.07\pm0.02$ for the hard one, and
agrees with the Belle index $n=10.5\pm1.2\pm0.5$ that the hard result
could not reproduce in Sec.~\ref{subsec:hardresults}.  For $\etaeta$, whose
hard index was already compatible with the data, the coherent index
$8.3\pm0.4$ between $2.4$ and $3.3~\mathrm{GeV}$ agrees with the measured
$7.8\pm0.6\pm0.4$ as well.  The normalization stays below the data, as
for $\pizpiz$ and $\KsKs$~\cite{He:2026bfv}.  Higher-order corrections
may account for part of this gap, since they can strongly affect the
absolute cross sections.  For the charged pairs the next-to-leading-order correction
to the collinear twist-2 cross section reaches 90\% of the leading-order
one at $Q=4~\mathrm{GeV}$~\cite{Duplancic:2006nv}.

\section{Summary}
\label{sec:summary}

We have computed the two-photon production of the pseudoscalar pairs
$\pieta$, $\pietap$, $\etaeta$, $\etaetap$, and $\etapetap$ between $2$ and
$4~\mathrm{GeV}$ in $k_{T}$ factorization, extending the treatment of
$\pizpiz$ and $\KsKs$~\cite{He:2026bfv} to final states with a
flavor-singlet component.  The hard amplitudes contain the twist-2 quark
and two-gluon contributions and the chirally enhanced twist-3
contributions of the nonstrange and strange flavor states, and the soft
handbag amplitude is added coherently to them.  The two-gluon component
and the soft contribution, neither of which had been combined with the
twist-3 hard amplitudes in these channels, are thus treated together with
them.

The hard cross sections are dominated by the twist-3 sector in all five
channels.  Its chiral enhancement $\kappa_{2}/\kappa_{1}$ falls only from
about $7$ to about $2$ between $2$ and $4~\mathrm{GeV}$, and the twist-3
amplitude keeps an end point contribution that survives the Sudakov and
threshold suppression.  The cross sections rise by about an order of
magnitude over twist 2, while the effective power grows only from about
$n=6.8$ to $7.2$, so the enhancement persists over the whole range.  This
extends to the $\etapp$ channels the twist-3 dominance found for $\pizpiz$
and $\KsKs$.  The comparison with Belle separates the two measured
channels.  For $\etaeta$ the twist-3 result comes close to the data, with
$n=7.31\pm0.09$ against the measured $7.8\pm0.6\pm0.4$, whereas for $\pieta$ it
lies 3 to 5 times below the data, which fall with
$n=10.5\pm1.2\pm0.5$ against $7.07\pm0.02$ for the twist-3 result.

The singlet component sorts the channels by their sensitivity to the
two-gluon content.  The singlet coupling of the $\eta$ is small,
$f_{\eta}^{1}=25~\mathrm{MeV}$, because its nonstrange and strange parts
nearly cancel, so varying the two-gluon LCDA within its uncertainty changes the cross
sections of the measured channels by about 2\% and these channels test the quark sector.  The $\etap$ coupling
is 6 times larger, and the variation reaches 13\%, 20\%, and 24\%
in $\pietap$, $\etaetap$, and $\etapetap$ at $3~\mathrm{GeV}$.  In
$\etaetap$ the strange quark amplitude cancels more than half of the
nonstrange one, which makes this channel the smallest and the most
sensitive to the mixing parameters, and at twist 2 also to the two-gluon
content.

The energy dependence of the cross section ratios to $\pizpiz$ is the result that
discriminates between the mechanisms.  Each mechanism alone falls with
energy in the same way in all channels and gives nearly constant ratios,
$0.48$ and $0.77$ for the hard one and $0.43$ and $0.58$ for the handbag
in $\pieta$ and $\etaeta$.  Below $3.2~\mathrm{GeV}$ the measured $\pieta$
ratio lies above both values and the $\etaeta$ ratio below them, which no incoherent
sum can reach.  With the handbag amplitude added coherently, its modulus
taken from the pion and kaon channels and its phase fitted separately for
the isovector and the isoscalar pairs, the $\pieta$ ratio is reproduced
from $2.4$ to $4~\mathrm{GeV}$, the interference turning from constructive to destructive as the
phase rotates by about $84^{\circ}$ per GeV.  In $\etaeta$ the
interference is destructive and lowers the ratio from the hard value $0.77$
to about $0.36$ below $2.8~\mathrm{GeV}$, in agreement with the data, while the highest points,
close to the $\chi_{c0}$ resonance, fall below the coherent result.  The
same interference steepens the $\pieta$ cross section to $n=9.7\pm1.6$ and so
accounts for the energy dependence that the hard result missed.  As for
$\pizpiz$ and $\KsKs$, the energy dependence of the ratios can thus be
understood in the coherent description as the interference of two
mechanisms of comparable size whose relative phase rotates with energy.  For $\pietap$ the soft amplitude stands to that of $\pieta$ in the
same ratio $\tan\phi$ as the hard quark amplitudes, so the predicted ratio
of the two channels stays close to $\tan^{2}\phi=0.67$ and tests the mixing
scheme nearly independently of the soft parameters.

Two limits remain.  The soft
phase is fitted rather than computed, so a good fit does not by itself
establish the handbag as the soft mechanism.  The isoscalar phase in
particular is determined by the $\etaeta$ ratio alone, measured only up
to $3.3~\mathrm{GeV}$, and is carried over to $\etaetap$ and $\etapetap$
without an independent test.  A measurement of $\etaetap$ or $\etapetap$, or of $\etaeta$
above the charmonium region, would constrain it, and in a joint analysis
the $\etap$ channels could also help determine the sign of the two-gluon
coefficient
$a_{2}^{g}$, which the transition form factors leave open.  Finally, the
normalization of the cross sections stays below the data even with the
soft term, as for $\pizpiz$ and $\KsKs$.  Higher-order corrections can be
large at these energies, the next-to-leading-order term of the collinear twist-2
cross section of the charged pairs reaching 90\% of the
leading-order one at $4~\mathrm{GeV}$~\cite{Duplancic:2006nv}, and may
account for part of the remaining gap.  A quantitative account of the
normalization therefore depends on the next-to-leading-order corrections
to the twist-3 amplitudes, which are not yet available.

\begin{acknowledgments}
This work was supported by the National Natural Science Foundation of
China under Grant No.~12305086 and the Project of Science and Technology
Research Program of the Hubei Provincial Department of Education under
Grant No.~Q20222504.
\end{acknowledgments}

\appendix

\section{Hard kernels and resummation factors}
\label{app:hard}

This appendix collects the ingredients of Eq.~\eqref{eq:Aq} that are
common to the pion and kaon channels~\cite{Wang:2019trq,He:2026bfv}, in
the form used in the present calculation.

The oscillator parameter of Eq.~\eqref{eq:bspaceSigma} follows from the
mean intrinsic transverse momentum of the flavor state
\begin{equation}
\label{eq:kperpaverage}
\langle\kperp^{2}\rangle_{r}
=\frac{1}{2\beta_{r}^{2}}
\frac{\int_{0}^{1}\mathrm{d}x\,[\phi_{r}^{\,\pi}(x,\mu)]^{2}}
     {\int_{0}^{1}\mathrm{d}x\,[\phi_{r}^{\,\pi}(x,\mu)]^{2}/(x\xbar)}.
\end{equation}
The $x\xbar$ in the exponent of Eq.~\eqref{eq:bspaceSigma} narrows the
transverse distribution toward the end points, a correlation between the
longitudinal and the transverse degrees of freedom that a factorized
Gaussian would miss.

The strong coupling is evaluated at one loop with $\alpha_{s}(M_{Z})=0.1180$
and matched across the charm and bottom thresholds at $1.27$ and
$4.18~\mathrm{GeV}$, which gives $\LamQCD^{(4)}\simeq0.12~\mathrm{GeV}$.
The $b$ integration is cut at $b_{\max}=4.5~\mathrm{GeV}^{-1}$, below
the Landau pole, where the Sudakov factor has already suppressed the
integrand.

In the next-to-leading-logarithm approximation the Sudakov exponent
reads~\cite{Botts:1989kf,Li:1992nu,Coriano:1998ge}
\begin{equation}
\label{eq:SudakovS}
S(x,y,Q,b,t)=s(x,b,Q)+s(\xbar,b,Q)+s(y,b,Q)+s(\ybar,b,Q)
-\frac{2}{\bar\beta_{0}}\ln\frac{\ln(t/\LamQCD)}{\ln[1/(b\LamQCD)]},
\end{equation}
where the last term is the renormalization-group evolution between $1/b$
and the hard scale $t$, and $\exp[-S]$ is set to unity wherever $S<0$.
With $\hat q=\ln[\xi Q/(\sqrt2\LamQCD)]$ and
$\hat b=\ln[1/(b\LamQCD)]$, the function $s(\xi,b,Q)$ vanishes for
$b\le\sqrt2/(\xi Q)$ and is otherwise given by
\begin{equation}
\label{eq:sfun}
\begin{aligned}
s(\xi,b,Q)=\;&\frac{A^{(1)}}{2\bar\beta_{0}}\,\hat q\ln\frac{\hat q}{\hat b}
+\frac{A^{(2)}}{4\bar\beta_{0}^{2}}\left(\frac{\hat q}{\hat b}-1\right)
-\frac{A^{(1)}}{2\bar\beta_{0}}\bigl(\hat q-\hat b\bigr)\\
&-\frac{A^{(1)}\bar\beta_{1}}{4\bar\beta_{0}^{3}}\,\hat q
 \left[\frac{\ln(2\hat b)+1}{\hat b}-\frac{\ln(2\hat q)+1}{\hat q}\right]
-\left[\frac{A^{(2)}}{4\bar\beta_{0}^{2}}
 -\frac{A^{(1)}}{4\bar\beta_{0}}\ln\frac{e^{2\gamma_{E}-1}}{2}\right]
 \ln\frac{\hat q}{\hat b}\\
&+\frac{A^{(1)}\bar\beta_{1}}{8\bar\beta_{0}^{3}}
 \left[\ln^{2}(2\hat q)-\ln^{2}(2\hat b)\right],
\end{aligned}
\end{equation}
with $\bar\beta_{0}=(33-2n_{f})/12$, $\bar\beta_{1}=(153-19n_{f})/24$,
$A^{(1)}=4/3$, and
$A^{(2)}=67/9-\pi^{2}/3-10n_{f}/27+(8/3)\bar\beta_{0}\ln(e^{\gamma_{E}}/2)$,
evaluated at $n_{f}=4$ and with $\gamma_{E}$ the Euler constant.

The threshold logarithms are resummed into the jet
function~\cite{Li:1998is,Kurimoto:2001zj,Li:2001ay}
\begin{equation}
\label{eq:St}
S_{t}(x,Q)=\frac{2^{1+2c}\,\Gamma(\tfrac32+c)}{\sqrt\pi\,\Gamma(1+c)}
\bigl[x\xbar\bigr]^{c},
\end{equation}
whose exponent is taken from the fit to the pion transition form factor
in Ref.~\cite{Li:2009pr}
\begin{equation}
\label{eq:cQ}
c(Q)=\min\bigl\{0.04\,Q^{2}-0.51\,Q+1.87,\,1\bigr\},
\end{equation}
with $Q$ in GeV.  Since $S_{t}$ vanishes at $x\to0,1$, it damps the
end point region in which the twist-3 LCDA $\phi_{r}^{p}\simeq1$ would
otherwise enhance the convolution.

The virtualities of the internal gluon in the four groups are
\begin{equation}
\label{eq:sn}
\begin{aligned}
s_{a}&=\xbar y\,Q^{2},
&\qquad s_{b}&=x\ybar\,Q^{2},\\
s_{c}&=(x-c^{2})(\ybar-s^{2})\,Q^{2},
&\qquad s_{d}&=(\xbar-c^{2})(y-s^{2})\,Q^{2},
\end{aligned}
\end{equation}
and the hard scales are
\begin{equation}
\label{eq:tn}
\begin{aligned}
t_{a}&=\max\bigl\{Q\sqrt{\xbar y},\,1/b\bigr\},
&\qquad t_{b}&=\max\bigl\{Q\sqrt{x\ybar},\,1/b\bigr\},\\
t_{c}&=\max\bigl\{Q\sqrt{c^{2}\xbar\ybar+s^{2}xy},\,1/b\bigr\},
&\qquad t_{d}&=\max\bigl\{Q\sqrt{s^{2}\xbar\ybar+c^{2}xy},\,1/b\bigr\}.
\end{aligned}
\end{equation}
The virtualities $s_{a}$ and $s_{b}$ are positive throughout the physical
region, whereas $s_{c}$ and $s_{d}$ change sign, so the Fourier
transformed gluon propagator takes the two forms
\begin{equation}
\label{eq:Fsnb}
\mathrm{F}(s_{n},b)=
\begin{cases}
2\pi\,\Hone(\sqrt{s_{n}}\,b), & s_{n}>0,\\[3pt]
-4i\,\Kzero(\sqrt{-s_{n}}\,b), & s_{n}<0,
\end{cases}
\end{equation}
with $\Hone$ the Hankel function of the first kind and $\Kzero$ the
modified Bessel function of the second kind.  The groups $c$ and $d$ carry
in addition the Bessel functions $\Jzero(pb)$, $\Jone(pb)$, and $\Jtwo(pb)$ of the photon-side transverse momentum $p=Qsc$.

The kernels carry two labels $i,j\in\{\pi,p,\sigma,\sigma'\}$, the first
for the LCDA of $P_{2}$ and the second for that of $P_{1}$.  Mixed
twist-2 and twist-3 kernels vanish, since a single chirality flip cannot
be compensated in the chiral limit.  Besides the kernels of
Eqs.~\eqref{eq:Ta2} and~\eqref{eq:Ta3}, group $a$ contributes
\begin{align}
\label{eq:TaApp}
\TildeT^{++}_{ap\sigma'}&=\frac{1}{27}\kappa_{2}
\left[\frac{1}{s^{2}c^{2}}-\frac{2}{x}\right]\mathrm{F}(s_{a},b),
\nonumber\\
\TildeT^{+-}_{ap\sigma'}&=\frac{1}{27}\kappa_{2}
\left[\frac{\xbar}{s^{2}c^{2}x}-\frac{2}{x\xbar}\right]\mathrm{F}(s_{a},b),
\nonumber\\
\TildeT^{++}_{a\sigma'\sigma'}&=-\frac{1}{162}\kappa_{2}
\left[\frac{s^{2}(1-\xbar y)}{c^{2}x\ybar}
     +\frac{c^{2}(1-\xbar y)}{s^{2}x\ybar}\right]\mathrm{F}(s_{a},b),
\nonumber\\
\TildeT^{+-}_{a\sigma'\sigma'}&=-\frac{1}{162}\kappa_{2}
\left[\frac{s^{2}(1+\xbar y)}{c^{2}x\ybar}
     +\frac{c^{2}(1+\xbar y)}{s^{2}x\ybar}\right]\mathrm{F}(s_{a},b).
\end{align}
The independent kernels of group $c$ are, at twist 2,
\begin{align}
\label{eq:Tc2}
\TildeT^{++}_{c\pi\pi}&=\frac{4}{9}\kappa_{1}
\left[\frac{xy+\xbar\ybar}{s^{2}xy}+\frac{xy+\xbar\ybar}{c^{2}\xbar\ybar}\right]
\Jzero(pb)\,\mathrm{F}(s_{c},b),
\nonumber\\
\TildeT^{+-}_{c\pi\pi}&=-\frac{4}{9}\kappa_{1}
\left[4-\frac{s^{2}(\xbar y+x\ybar)}{c^{2}\xbar\ybar}
       -\frac{c^{2}(\xbar y+x\ybar)}{s^{2}xy}\right]
\Jzero(pb)\,\mathrm{F}(s_{c},b),
\end{align}
and at twist 3
\begin{align}
\label{eq:Tc3}
\TildeT^{++}_{cpp}&=-\frac{2}{9}\kappa_{2}
\left[2+\frac{1-\xbar\ybar}{s^{2}xy}+\frac{1-xy}{c^{2}\xbar\ybar}\right]
\Jzero(pb)\,\mathrm{F}(s_{c},b),
\nonumber\\
\TildeT^{+-}_{cpp}&=-\frac{2}{9}\kappa_{2}
\left[\frac{1}{s^{2}c^{2}}-2\right]\Jzero(pb)\,\mathrm{F}(s_{c},b),
\nonumber\\
\TildeT^{++}_{cp\sigma'}&=\frac{1}{27}\kappa_{2}
\left[\frac{s^{2}-c^{2}}{s^{2}c^{2}}-\frac{x-\xbar}{x\xbar}\right]
\Jzero(pb)\,\mathrm{F}(s_{c},b),
\nonumber\\
\TildeT^{+-}_{cp\sigma'}&=\frac{1}{27}\kappa_{2}
\left[\frac{s^{2}x^{2}-c^{2}\xbar^{2}}{s^{2}c^{2}x\xbar}
     +\frac{x-\xbar}{x\xbar}\right]\Jzero(pb)\,\mathrm{F}(s_{c},b),
\nonumber\\
\TildeT^{++}_{cp\sigma}&=-\frac{1}{27}\kappa_{2}\,
\frac{\omega b\,(s^{2}x+c^{2}\xbar)}{sc\,x\xbar}\,\Jone(pb)\,\mathrm{F}(s_{c},b),
\nonumber\\
\TildeT^{+-}_{cp\sigma}&=-\frac{1}{27}\kappa_{2}\,
\frac{\omega b\,(c^{2}x+s^{2}\xbar+1)}{sc\,x\xbar}\,\Jone(pb)\,\mathrm{F}(s_{c},b),
\nonumber\\
\TildeT^{++}_{c\sigma'\sigma'}&=-\frac{1}{162}\kappa_{2}
\left[\frac{s^{2}(1-xy)}{c^{2}\xbar\ybar}+\frac{c^{2}(1-\xbar\ybar)}{s^{2}xy}\right]
\Jzero(pb)\,\mathrm{F}(s_{c},b),
\nonumber\\
\TildeT^{+-}_{c\sigma'\sigma'}&=-\frac{1}{162}\kappa_{2}
\left[\frac{s^{2}(1+xy)}{c^{2}\xbar\ybar}+\frac{c^{2}(1+\xbar\ybar)}{s^{2}xy}\right]
\Jzero(pb)\,\mathrm{F}(s_{c},b),
\nonumber\\
\TildeT^{++}_{c\sigma'\sigma}&=\frac{1}{162}\kappa_{2}\,\omega b
\left[\frac{s}{c\,\xbar}-\frac{c}{s\,x}\right]\Jone(pb)\,\mathrm{F}(s_{c},b),
\nonumber\\
\TildeT^{+-}_{c\sigma'\sigma}&=\frac{1}{162}\kappa_{2}\,\omega b
\left[\frac{s(1+y)}{c\,\xbar\ybar}-\frac{c(1+\ybar)}{s\,xy}\right]
\Jone(pb)\,\mathrm{F}(s_{c},b),
\nonumber\\
\TildeT^{+-}_{c\sigma\sigma}&=-\frac{1}{81}\kappa_{2}\,\omega^{2}b^{2}
\left[\frac{1}{xy}+\frac{1}{\xbar\ybar}\right]\Jtwo(pb)\,\mathrm{F}(s_{c},b).
\end{align}
The tensor-tensor kernel feeds $\mathcal{M}_{+-}$ alone, its
equal-helicity projection vanishing identically, and the kernels with an undifferentiated $\phi_{r}^{\sigma}$ appear only in groups
$c$ and $d$, through the $\partial/\partial k_{\perp\nu}$ term of
Eq.~\eqref{eq:projector} acting on the photon-side transverse momentum.

All other kernels follow from discrete symmetries of the partonic
amplitude.  Exchanging the two LCDA labels within a group corresponds to
$x\to\ybar$, $y\to\xbar$, and $s\leftrightarrow c$
\begin{equation}
\label{eq:Tswap}
\TildeT^{\lambda_{1}\lambda_{2}}_{nji}
=\epsilon_{ij}\,\TildeT^{\lambda_{1}\lambda_{2}}_{nij}
\Big|_{x\to\ybar,\;y\to\xbar,\;s\leftrightarrow c},
\qquad n=a,c,
\end{equation}
with $\epsilon_{ij}=-1$ for $(i,j)=(p,\sigma')$ and $(\sigma',\sigma)$ and
$\epsilon_{ij}=+1$ for $(p,\sigma)$.  Group $b$ follows from group $a$ by
the exchange of the two mesons, $\theta^{*}\to\pi-\theta^{*}$,
\begin{equation}
\label{eq:Tab}
\TildeT^{\lambda_{1}\lambda_{2}}_{bij}
=\TildeT^{\lambda_{1}\lambda_{2}}_{aji}
\Big|_{x\leftrightarrow y,\;s\leftrightarrow c},
\end{equation}
and group $d$ from group $c$ by the interchange of quark and antiquark
inside both mesons
\begin{equation}
\label{eq:Tcd}
\TildeT^{\lambda_{1}\lambda_{2}}_{dij}
=\eta_{ij}\,\TildeT^{\lambda_{1}\lambda_{2}}_{cij}
\Big|_{x\to\xbar,\;y\to\ybar},
\end{equation}
with $\eta_{ij}=-1$ for the pairs containing exactly one $\sigma'$ together
with $p$ or $\sigma$, that is $(p,\sigma')$, $(\sigma',p)$, $(\sigma',\sigma)$, and $(\sigma,\sigma')$, and $\eta_{ij}=+1$ otherwise.  The sign reflects
the parity of the derivative $\phi_{r}^{\sigma'}$ under
$x\leftrightarrow\xbar$.  Together these relations generate all $58$
kernels of the four groups, counted separately for the two photon
helicities, from the $21$ given explicitly, which also serves as an
internal check of the calculation.  The twist-2 kernels can also be
checked against the collinear limit.  Taking $\kperp\to0$ amounts to
integrating the kernels over $b$ with $b$-independent wave functions,
which turns $\mathrm{F}(s_{n},b)$ into the propagator $1/s_{n}$ in groups
$a$ and $b$, while the factor $\Jzero(pb)$ shifts the virtuality of
groups $c$ and $d$ to $s_{n}-p^{2}$.  Summed over the four groups, the
twist-2 kernels then reproduce the leading-twist collinear amplitudes of
Ref.~\cite{Brodsky:1981rp} for a pair of neutral mesons.

\section{Flavor relations of the annihilation form factors}
\label{app:soft}

We extend the $SU(3)$ analysis of Ref.~\cite{Diehl:2009yi} from the
octet state $\eta_{8}$ to the flavor states $\eta_{q}$ and $\eta_{s}$,
writing it in terms of flavor invariants.
Each meson of the present channels is flavor neutral,
$P=\sum_{q}a_{q}^{P}\,q\bar q$ with the flavor coefficients $a_{q}^{P}$ listed above
Eq.~\eqref{eq:Rrule}, and is represented by the flavor-diagonal matrix
$\Phi_{P}=\mathrm{diag}(a_{u}^{P},a_{d}^{P},a_{s}^{P})$.  The form factor
$R^{q}_{P_{1}P_{2}}$ is the matrix element of a source that creates the
flavor $q\bar q$, represented by the flavor matrix $\Pi_{q}$ whose only
nonzero entry is a unit at position $qq$.  Under a flavor rotation $V$
the source and meson matrices all transform as
$\Phi\to V\Phi V^{\dagger}$, and $SU(3)$ symmetry requires the form
factor to be built from the invariants of $\Pi_{q}$, $\Phi_{P_{1}}$, and $\Phi_{P_{2}}$, that is from traces of their products.  For octet mesons
two invariants contribute.  In the first the quark line of the source
runs through both mesons, its quark entering one and its antiquark the
other, which gives
$\mathrm{Tr}\,\Pi_{q}\Phi_{P_{1}}\Phi_{P_{2}}=a_{q}^{P_{1}}a_{q}^{P_{2}}$.
In the second the source line closes on itself and the pair is formed by
a separate quark line, which gives the product of $\mathrm{Tr}\,\Pi_{q}=1$
and the flavor overlap
$\mathrm{Tr}\,\Phi_{P_{1}}\Phi_{P_{2}}=\sum_{q'}a_{q'}^{P_{1}}a_{q'}^{P_{2}}$,
independent of $q$.  The form factor is a combination of these two
invariants, whose weights are fixed by the neutral pion pair.  The valence
and nonvalence form factors of Ref.~\cite{Diehl:2009yi} are defined as
the flavor form factors of $\pi^{+}\pi^{-}$,
$R^{u}_{2\pi}=R^{u}_{\pi^{+}\pi^{-}}$ and
$R^{s}_{2\pi}=R^{s}_{\pi^{+}\pi^{-}}$, and isospin symmetry makes those
of $\pizpiz$ equal to them.  Matching the two invariants to
$R^{u}_{\pizpiz}=R^{u}_{2\pi}$ and $R^{s}_{\pizpiz}=R^{s}_{2\pi}$ gives
\begin{equation}
\label{eq:RqGeneral}
R^{q}_{P_{1}P_{2}}
=R^{s}_{2\pi}\,\mathrm{Tr}\,\Phi_{P_{1}}\Phi_{P_{2}}
+2\bigl(R^{u}_{2\pi}-R^{s}_{2\pi}\bigr)\,a_{q}^{P_{1}}a_{q}^{P_{2}}.
\end{equation}
The overlap term, in which the source closes on itself, is OZI
suppressed and weighted by the nonvalence form factor, while the other
term, which carries the source quarks into the two mesons, alone
survives in the valence limit $R^{s}_{2\pi}\to0$.
For $\pi^{0}\eta_{8}$ and $\eta_{8}\eta_{8}$ this reproduces the relations
of Ref.~\cite{Diehl:2009yi}, for instance
$R^{u}_{\eta_{8}\pi^{0}}=(R^{u}_{2\pi}-R^{s}_{2\pi})/\sqrt3$ and
$R^{s}_{\eta_{8}\eta_{8}}=\tfrac43R^{u}_{2\pi}-\tfrac13R^{s}_{2\pi}$.

Once the singlet is included, $SU(3)$ admits two further invariants,
containing the flavor trace of one or of both mesons, in which such a
meson is attached to the rest of the diagram only through its flavor
trace, that is through its singlet component and gluons.  They vanish for
octet mesons and are therefore not constrained by the analysis of
Ref.~\cite{Diehl:2009yi}.  They violate the OZI rule in the same way as
the coupling of the $\eta_{s}$ to a nonstrange current that the
Feldmann-Kroll-Stech scheme neglects~\cite{Feldmann:1998vh}, and we drop
them, so that Eq.~\eqref{eq:RqGeneral} holds for the flavor states as well.

With $r_{s}=R^{s}_{2\pi}/R^{u}_{2\pi}$, Eqs.~\eqref{eq:Rsum}
and~\eqref{eq:RqGeneral} give for the five flavor-state pairs
\begin{align}
\label{eq:Rflavorrs}
\frac{R_{\pi^{0}\eta_{q}}}{R_{\pizpiz}}&=\frac{3(1-r_{s})}{5+r_{s}},
&
\frac{R_{\eta_{q}\eta_{q}}}{R_{\pizpiz}}&=1,
&
\frac{R_{\eta_{s}\eta_{s}}}{R_{\pizpiz}}&=\frac{2+4r_{s}}{5+r_{s}},
\nonumber\\
R_{\pi^{0}\eta_{s}}&=0,
&
R_{\eta_{q}\eta_{s}}&=0,
\end{align}
with $R_{\pizpiz}=\tfrac19(5+r_{s})R^{u}_{2\pi}$.  The two pairs without a
common valence flavor vanish at any $r_{s}$, because their flavor overlap
$\mathrm{Tr}\,\Phi_{P_{1}}\Phi_{P_{2}}$ is zero as well.  The ratio $\eta_{q}\eta_{q}$ to
$\pizpiz$ is independent of $r_{s}$, the two pairs having the same flavor
overlap and the same valence charges.  The nonvalence form factor
matters in the other two.  In $\pi^{0}\eta_{q}$ it enters through the
$u$ and $d$ quarks with opposite signs, which removes the overlap term
and leaves $R^{u}_{2\pi}-R^{s}_{2\pi}$.  In $\eta_{s}\eta_{s}$ the valence
charge is the small $e_{s}^{2}$, while the nonvalence term couples to the
$u$ and $d$ quarks with $e_{u}^{2}+e_{d}^{2}=5/9$ and carries 2 times the
weight of the valence term.  This is the same ratio $(2+4r_{s})/(5+r_{s})$
that relates $R_{K^{0}\bar K^{0}}$ to $R_{K^{+}K^{-}}$~\cite{Diehl:2009yi},
and for the same reason, so that the strongest sensitivity to $r_{s}$
appears in channels with a weakly charged valence quark.

For the reasons given in Sec.~\ref{subsec:soft} we work at $r_{s}=0$, the limit also
used for $R_{K^{0}\bar K^{0}}=\tfrac25R_{K^{+}K^{-}}$ in
Ref.~\cite{He:2026bfv}.  Equation~\eqref{eq:RqGeneral} then reduces to
Eq.~\eqref{eq:Rrule}, and Eq.~\eqref{eq:Rflavorrs} to
Eq.~\eqref{eq:Rflavor}.  In this limit the soft form factors follow the
same rule as the hard quark amplitude, the flavor entering only through
the charge-weighted overlap of the valence wave functions.

\bibliography{refs}

\end{document}